\documentclass[manuscript=article]{achemso}
\usepackage{caption}                                    % figure captions
\usepackage[colorlinks=true, allcolors=blue]{hyperref}  % make links blue
\usepackage{amssymb} %Array of math symbols and fonts
\usepackage{mathtools} %Extension of amsmath, fixes deficiencies
\usepackage[version=4]{mhchem}
\usepackage{siunitx}

\SectionNumbersOn
\newcommand*{\citen}[1]{%   %adds in-line Ref.
  \begingroup
    \romannumeral-`\x
    \setcitestyle{numbers}%
    \cite{#1}%
  \endgroup
}
\DeclareSIUnit{\angstrom}{\textup{\AA}}
\newcommand{\gexc}{g_{\mathrm{exc}}}
\newcommand{\Gads}{G_{\mathrm{ads}}}
\newcommand{\phiE}{\phi_{\mathrm{E}}}

\title{\textit{Ab-Initio}-Based Modeling of Thermodynamic Cyclic Voltammograms:\\
A Benchmark Study on Ag(100) in Bromide Solutions}

\author{Nicolas Bergmann}
\affiliation{Fritz-Haber-Institut der Max-Planck-Gesellschaft, Faradayweg 4-6, D-14195 Berlin, Germany}
\author{Nicolas G. H\"{o}rmann}
\affiliation{Fritz-Haber-Institut der Max-Planck-Gesellschaft, Faradayweg 4-6, D-14195 Berlin, Germany}
\email{hoermann@fhi-berlin.mpg.de}
\author{Karsten Reuter}
\affiliation{Fritz-Haber-Institut der Max-Planck-Gesellschaft, Faradayweg 4-6, D-14195 Berlin, Germany}

\date{\today}

\begin{document}

\begin{abstract}
Experimental cyclic voltammograms (CVs) measured in the slow scan rate limit can be entirely described in terms of thermodynamic equilibrium quantities of the electrified solid-liquid interface. They correspondingly serve as an important benchmark for the quality of first-principles calculations of the interfacial thermodynamics. Here, we investigate the partially drastic approximations made presently in computationally efficient such calculations for the well-defined showcase of a Ag(100) model electrode in Br-containing electrolytes, where the non-trivial part of the CV stems from the electrosorption of Br ions. We specifically study the entanglement of common approximations in the treatment of solvation and field effects, as well as in the way macroscopic averages of the two key quantities, namely the potential-dependent adsorbate coverage and electrosorption valency, are derived from the first-principles energetics. We demonstrate that the combination of energetics obtained within an implicit solvation model and a perturbative second order account of capacitive double layer effects with a constant-potential grand-canonical Monte Carlo sampling of the adsorbate layer provides an accurate description of the experimental CV. However, our analysis also shows that error cancellation at lower levels of theory may equally lead to good descriptions even though key underlying physics like the disorder-order transition of the Br adlayer at increasing coverages is inadequately treated. 
\end{abstract}

\section{Introduction}
\label{sec:introduction}
Cyclic Voltammetry is a widely employed electrochemical experiment to characterize electrocatalytic processes occurring at electrified solid-liquid interfaces.\cite{bard2000voltammetry, nicholson1965theory} A cyclic voltammogram (CV) records the electric current $j$ observed while sweeping an applied electrode potential $\phiE$ at a constant scan rate $\nu$ upwards and downwards within a given potential window.\cite{elgrishi2018practical, kissinger1983cyclic, climent2018cyclic, engstfeld2018polycrystalline} Peaks in the resulting voltammogram $j(\phiE)$ are then interpreted as fingerprints of occurring electrochemical reactions,\cite{climent2018cyclic} whose fundamental nature can experimentally be uncovered by e.g. investigating the CV's dependencies on pH, applied potential limits, scan rate or the electrolyte's chemical composition.\cite{sheng2015correlating, aristov2015cyclic} The derived assignments are often not unambiguous though and could strongly benefit from independent and predictive-quality computational modeling.
\\
Whenever diffusion limitations are absent, CV currents $\left(j={\mathrm d}\sigma/{\mathrm d}t \right)$ within the stable potential window of the electrolyte directly relate to changes in the equilibrium electronic surface charge ${\mathrm d}\sigma$, due to differential electrode potential changes ${\mathrm d}\phiE$ induced by the constant scan rate $\nu = {\mathrm d}\phiE/{\mathrm d}t$. In this case, the dominant charging processes are the polarization of the electrolyte solution via double layer (DL) charging (${\mathrm d}\sigma_{\rm DL}$) and Faradaic processes in which charged particles transfer across the electrode. In the defined case of a stable model electrode surface that neither reconstructs nor dissolves, on which we will focus here, the electronic charge transfer (${\mathrm d}\sigma_{\rm a}$) due to the latter processes stems entirely from electrosorption of adsorbates ${\rm a}$ onto the surface.\cite{climent2018cyclic, karlberg2007cyclic, hoermann2021thermodynamic} ${\mathrm d}\sigma_{\rm a}$ is then given by the change in adsorbate coverage $\theta_{\rm a}$ multiplied by the number of exchanged electrons per adsorbate, aka the electrosorption valency $l_{\mathrm{a}}$.\cite{schultze1973experimental, guidelli2005electrosorption} One can thus formally write
\begin{eqnarray}
    j(\phiE) = \nu 
        \frac{{\mathrm{d}}\sigma_{\mathrm{DL}}}
        {{\mathrm{d}}\phiE} 
        + \nu 
        l_{\mathrm{a}}\left(\theta_{\mathrm{a}}, \phiE\right)
        \frac{{\mathrm{d}} \theta_{\mathrm{a}}}
        {{\mathrm{d}}\phiE} 
    \ .
    \label{eq:CV}
\end{eqnarray}
With $\sigma_{\rm DL} (\phiE)$ often a quasi constant baseline current, the theoretical modeling of a CV correspondingly requires an accurate description of the coverage vs potential relation $\theta_{\rm a}(\phiE)$, as well as an appropriate consideration of the electrosorption valency $l_{\mathrm{a}}\left(\theta_{\mathrm{a}}, \phiE\right)$. In the slow scan rate limit, both of these quantities are thermodynamic equilibrium quantities. In this respect, corresponding experimental CVs also serve as important benchmarks for the quality of theoretical predictions of the interfacial thermodynamics.
\\
It is with this motivation to benchmark various prevalent thermodynamic modeling choices and approximations that we here study the CV of a Ag(100) electrode in \ce{Br-}-containing electrolyte. This is a suitable and experimentally well-studied prototype system,\cite{wandlowski2001adsorption, koper1998_lattice_gas, nakamura2011structure} for which high-quality CVs are available and in which electronic charge transfer arises from the electrosorption of \ce{Br-} ions onto defined high-symmetry sites of an otherwise rigid Ag(100) lattice. We specifically compare popular choices made in three significant modeling steps: The modeling of the liquid-solid interface, the determination of the energetics at applied potential conditions, and the statistical mechanics description to obtain macroscopic averages of $l_{\mathrm{a}}$ and $\theta_{\mathrm{a}}$ from the atomistic energetics. The focus is thereby on computationally efficient approaches based on density-functional theory (DFT) calculations using a slab model for the electrode and without explicit representation of the electrolyte solution. We thus compare vacuum calculations to those in an implicit solvent environment, consider applied potential effects in first- and second-order\cite{norskov2004origin, hormann2020electrosorption}, and apply mean-field and lattice-based grand-canonical Monte Carlo (GC-MC) sampling\cite{hoermann2021thermodynamic, koper1998_lattice_gas}. The analysis shows that only higher-order thermodynamics coupled to the lattice GC-MC sampling consistently recreates the characteristic peak shape and integral of the experimental CVs for the right reasons.

\section{Experimental CVs of Ag(100) in Bromide Solutions}
\label{sec:experimental_CVs}

\begin{figure}[htpb]
    \centering
    \includegraphics[width=1\textwidth]{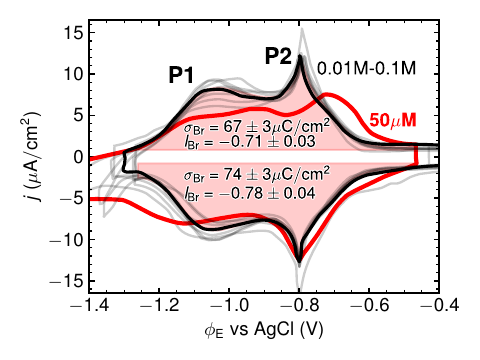}
    \caption{10 experimental CVs of Ag(100) in \ce{Br-}-containing electrolytes (gray lines) obtained from the echemdb database\cite{engstfeld2023echemdb}, with $\nu$- and $c_{\ce{Br-}}$-normalized to \SI{50}{mV/s} and \SI{0.1}{M}, respectively. The CV measured by Nakamura {\em et al.}\cite{nakamura2011structure}	(bold black line) is henceforth taken as representative experimental reference in all figures below. In the anodic sweep direction, P1 marks the CV peak corresponding to the onset of Br electrosorption. At P2, the second-order disorder-order phase transition occurs to the final $c(2\times2)$ Br-covered surface. Integrating the CV over the potential range from [\SI{-1.3}{V}, \SI{-0.4}{V}] vs AgCl (indicated by the red shaded area) and subtracting the capacitive DL baseline current contribution yields the total transferred electrosorption charge $\sigma_{\rm Br}$ and derived from it the electrosorption valency $l_{\ce{Br}}$ in the anodic and the cathodic sweep direction. The values quoted in the figure correspond to the average and standard deviation over the 10 CVs. The 10 CVs are from measurements at higher electrolyte concentrations (0.01 M - 0.1 M), where thermodynamic CVs can be obtained at the applied scan rates. This is contrasted by the CV shown as a red line that was measured at $c_{\ce{Br-}}=\SI{50}{\mu M}$.\cite{endo1999situ} In this curve, the \ce{Br-} electrosorption's kinetic limitations become visible through the peak hysteresis between the anodic and cathodic sweep directions. See the SI for the individual CVs, their references, as well as all details regarding the CV normalization and integration.}
    \label{fig:experimental_Br_CVs}
\end{figure}

The model system Ag(100) in \ce{Br-}-containing electrolytes and its CV has been studied extensively.\cite{nakamura2011structure, ocko1997bromide, wandlowski2001adsorption, koper1998_lattice_gas, koper1998_MC_simulations, wang2002ab, mitchell2000dynamics, mitchell2002halide} Figure \ref{fig:experimental_Br_CVs} shows a collection of digitized experimental CVs, using the echemdb database\cite{engstfeld2023echemdb}. The CVs span a range of electrolyte concentrations and cations. CVs with concentrations above \SI{10}{mM} exhibit no significant hysteresis, i.e., the peak positions in the anodic sweep direction are essentially identical to those of the cathodic sweep direction, indicating the thermodynamic character of the experiments.\cite{climent2018cyclic} Kinetic limitations, likely due to \ce{Br-} diffusion,\cite{endo1999situ} only become relevant at much smaller concentrations (bold, red curve in Fig. \ref{fig:experimental_Br_CVs}), which will not be studied here.
\\
In general, the "butterfly"-shape of the CV is characterized by a first shoulder (peak P1) at lower potentials, which is ascribed to the formation of a disordered Br-adlayer with $\theta_{\rm Br} \leq$\,0.3~monolayer~(ML), as evidenced in surface X-ray scattering experiments by Wandlowski \emph{et al.}\cite{wandlowski2001adsorption}. The prominent sharp peak (P2) at $\approx$\SI{0.38}{ML}, i.e. $\sim 75\%$ of the limiting coverage \SI{0.5}{ML}, marks the second-order disorder-order phase transition where phase boundaries between different sub-lattices of Br adlayers are continuously removed,\cite{persson1992ordered, landau2013statistical} ultimately resulting in an ordered $\mathrm{c(2\times2)}$ Br adlayer with \SI{0.5}{ML} coverage at high potentials\cite{ocko1997bromide, wandlowski2001adsorption, koper1998_lattice_gas}.
\\
The total transferred electronic charge $\sigma_{\mathrm{Br}}$, as determined by integrating the CV without baseline currents (red shaded area in Fig. \ref{fig:experimental_Br_CVs}), indicates that the electrosorption valency is non-integer. Assuming a nominal full electron transfer during electrosorption of \ce{Br-}, i.e. $l_{\ce{Br}} = -1$, the expected transferred electronic charge to a \SI{0.5}{ML} adlayer would be $\sigma_{\mathrm{Br,~nominal}} = \SI{94}{\micro C\per\centi\meter\squared}$. This is 25\% higher than the actually measured value of $\sigma_{\mathrm{Br}} \approx \SI{70}{\micro C\per \centi\meter\squared}$, cf. Fig.~\ref{fig:experimental_Br_CVs}. Additionally, Ref. \citen{wandlowski2001adsorption} reported a non-Nernstian potential shift of the P2 peak of \SI{110}{\milli eV} per decadic logarithm of the \ce{Br-} concentration. Both of these observations are consistent with a non-integer electrosorption valency $l_{\ce{Br}} \sim -0.75$, where $l_{\ce{Br}}$ is defined as\cite{schultze1973experimental}
\begin{equation}
    l_{\ce{Br}} = -\frac{1}{e}\left(\frac{\partial \sigma_{\rm Br}}{\partial \theta_{{\rm Br}}}\right)_{\phi_{E}} =-\frac{1}{e} \frac{\left(\frac{\partial \theta_{{\rm Br}}}{\partial \phiE}\right)_{\tilde{\mu}_{\ce{Br-}}}}{ \left(\frac{\partial \theta_{\rm Br}}{\partial \tilde{\mu}_{\ce{Br-}}}\right)_{\phiE} } \quad , 
    \label{eq:electrosorption_valency}
\end{equation}
with $e$ the elementary charge and $\tilde{\mu}_{\ce{Br-}}$ the \ce{Br-} electrochemical potential. While not further discussed here, the known interdependencies of electrosorption valency and CV peak shapes \cite{hoermann2021peakpositions} indicate that this non-ideal $l_{\ce{Br}}$ value might as well explain the cation-dependence of the peak shape and integral observed in Ref. \citen{nakamura2011structure}.

\section{Theory}

\begin{figure}
    \centering
    \includegraphics[width=0.8\textwidth]{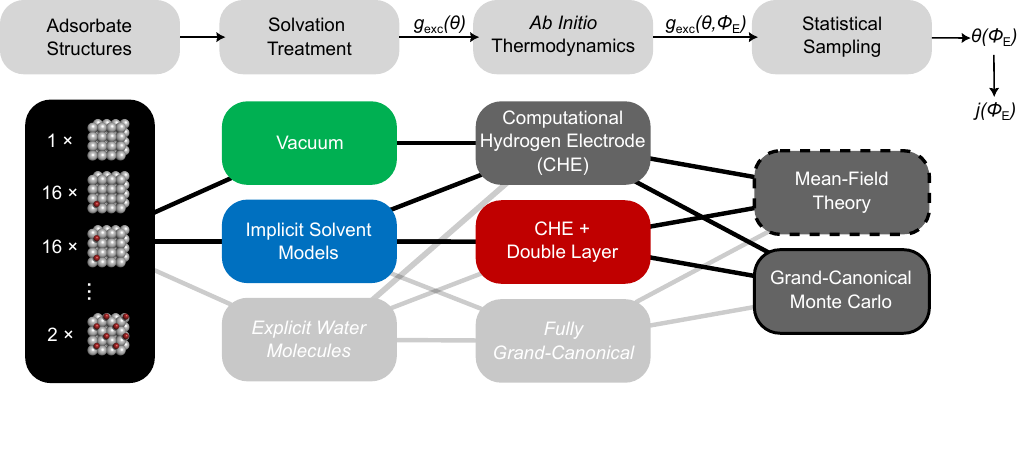}
    \caption{Typical modeling steps to derive thermodynamic CVs from first-principles calculations include the construction of a range of adsorbate structures on an electrode in a chosen solvent environment, the assessment of their stability using an \textit{ab initio} thermodynamics approach, and finally the determination of macroscopic averages via statistical sampling. The color codes and line forms of the different boxes reflect the colors used for corresponding data in all Figures below.}
    \label{fig:Introduction_alt}
\end{figure}

While computing CVs with effective models, e.g. mean-field or lattice Hamiltonians, based on fitted experimental parameters has a long tradition,\cite{koper1998_lattice_gas, mitchell2000dynamics, koper2000modeling} theoretical descriptions determined from first-principles methods based on DFT calculations are comparably recent\cite{norskov2004origin, karlberg2007cyclic}. A typical simulation workflow for such \textit{ab initio} thermodynamic CV modeling approaches is depicted in Fig. \ref{fig:Introduction_alt}. In the present work, we assess the impact of various choices at each of the indicated modeling steps. 
\\
In the first step, we determine the DFT energetics of a range of adsorbate structures in a given solvent environment, considering prevalent approximations of the latter in form of vacuum and an implicit solvent. Subsequently, we evaluate the relative stability of these structures as a function of electron and ion electrochemical potentials within an \textit{ab initio} thermodynamics framework. Finally, we perform thermodynamic averaging to obtain macroscopic averages of $\theta_{\rm Br}$ and $l_{\rm Br}$. Using eq.~\ref{eq:CV}, this then yields the computed CV curve, where we disregard the essentially structureless baseline current contribution $\sigma_{\rm DL}$ that was also approximately removed from the experimental CVs.

\subsection{Adsorbate Structures}

Br is experimentally known to adsorb onto the fourfold hollow sites of Ag(100) with a maximum coverage of 0.5\,ML in a regular $c(2 \times 2)$ arrangement as consistent with high nearest-neighbor (NN) repulsions.\cite{wang2002ab} We correspondingly construct a systematic dataset of adsorbate structures by enumerating all configurations in a $(4 \times 4)$ Ag(100) surface unit-cell that do not exhibit NN occupations\cite{koper1998_lattice_gas}. In total, this yields 28 symmetrically unique structures. To include information about the strong NN interactions in the dataset, we add a single 9/16\,ML structure with one additional Br on an empty site of the $c(2 \times 2)$ structure.

\subsection{Computational Method \& Solvation Treatment}

The energetics of all adsorbate structures is computed with DFT using the PBE functional\cite{perdew1996generalized} to treat electronic exchange and correlation. All calculations are performed with the Quantum ESPRESSO package\cite{giannozzi2009p, giannozzi2017advanced} and ultrasoft pseudopotentials from the GBRV database (GBRV 1.5)\cite{garrity2014pseudopotentials}, and are managed with the AiiDA-Quantum ESPRESSO pw workflow\cite{huber2022automated}. The $(4 \times 4)$ supercells employed to model the extended Ag(100) electrode comprise symmetric six layer slabs that are separated by a vacuum region of \SI{18.5}{\angstrom}.
For the implicit solvation (IS) we rely on the SCCS model \cite{andreussi2012revised, giannozzi2017advanced} with solvent parameters from H\"{o}rmann \emph{et al.}\cite{hoermann2019grand, hoermann2021thermodynamic} as implemented in the Quantum ENVIRON package \cite{andreussi2012revised}. Keeping the two innermost slab layers frozen at the optimized bulk distance, all structures are fully relaxed to energy and force thresholds below \SI{1.e-4}{Ry} and \SI{5.e-3}{Ry/Bohr}, respectively. Convergence tests indicate that at the employed computational settings (\texttt{ecutwfc}=\SI{45}{Ry}, \texttt{ecutrho}=\SI{360}{Ry} for the plane wave basis set, $(4 \times 4 \times 1)$ Gamma-centered $k$-point grid) the Br adsorption energies, $E_{\rm ads}$, are converged to within \SI{0.01}{eV}, with further details on the DFT calculations provided in the SI.
\\
With respect to the most relevant properties of the studied system, namely the work function and the adsorption energies, previous work indicates a better performance of the PBE functional as compared to other semi-local functionals.\cite{schimka2010accurate,schmidt2018benchmark} Nonetheless, PBE is known to underestimate formation energies of bulk halides by $\sim 0.4$ eV\cite{friedrich2019coordination, wang2021framework} and similar errors are reported for the adsorption energies of according species\cite{schmidt2018benchmark, wellendorff2015benchmark}. This generally needs to be kept in mind when judging predicted absolute CV peak positions and we will return to this point below. Fortunately, this uncertainty does not directly affect the here firstly aspired relative comparison of different computational approaches to the CV modeling that we consistently all base on the same PBE energetics.

\subsection{\textit{Ab Initio} Thermodynamics}

To evaluate the stability of adsorbate structures at applied electrode potential and experimental ion concentrations, we resort to two established electrochemical \textit{ab initio} thermodynamics approaches.\cite{reuter2016ab, scheffler1988thermodynamic, reuter2001composition, rogal2006ab, norskov2004origin, hoermann2021peakpositions}
\\
The most prominent method, the computational hydrogen electrode (CHE) approach\cite{norskov2004origin}, includes potential effects up to first order and proved successful in replicating experimental CV peaks for a variety of systems\cite{karlberg2007cyclic, tiwari2020electrochemical, tiwari2020fingerprint}. The CHE only necessitates the energetics at the potential of zero charge (PZC), without electronic excess charges on the metallic electrode, and can thus be evaluated in vacuum as well as in IS environments.\cite{gross2022ab, ringe2021implicit, dattila2022modeling, nitopi2019progress} Note, here and in previous works \cite{hormann2020electrosorption}, the term CHE refers only to its common application at PZC conditions, but not to its application at finite interfacial field or at finite, constant electronic excess charge. 
\\
In CHE, the stability of a structure $\alpha$ with $N^{\alpha}_{\rm Br}$ adsorbed Br atoms and $N^{\alpha}_{\mathrm{sites}}$ possible adsorption sites (and correspondingly a coverage $\theta_{\rm Br}^\alpha = N^{\alpha}_{\rm Br}/N^{\alpha}_{\mathrm{sites}}$) is given by the excess energy per surface site
\begin{equation}
    \gexc^{\alpha,\mathrm{CHE}} = \frac{1}{N^{\alpha}_{\mathrm{sites}}} 
        \left[
            G^{\alpha}_{\mathrm{surf},0} - G_{\ce{Ag},0}^{\mathrm{bulk}}
        \right] - \theta^{\alpha}_{\ce{Br}} \mu_{\ce{Br}} \ , 
    \label{eq:CHE_gexc}
\end{equation}
with $G$ here and henceforth referring to Gibbs free energies and the subscript $0$ to an evaluation at the PZC. $G^{\alpha}_{\mathrm{surf},0}$ ($G_{\ce{Ag},0}^{\mathrm{bulk}}$) is correspondingly the Gibbs free energy of the surface structure $\alpha$ (Ag bulk), and $\mu_{\mathrm{Br}}$ is the joint chemical potential for a charge-neutral Br species ($\ce{Br} = \ce{Br-} - \ce{e-}$) with
\begin{equation}
    \mu_{\mathrm{Br}}
        = \tilde{\mu}_{\ce{Br-}} - \tilde{\mu}_{e^{-}}
        =  \frac{1}{2}G_{\ce{Br2}(g)} + k_{\mathrm{B}}T\ln c_{\ce{Br-}} + e\left(\phiE - \phi_{\ce{Br}}^{\mathrm{ref}}\right) \quad .
    \label{eq:CHE_chemical_potential}
\end{equation}
Here, $G_{\ce{Br2}(g)}$ is the Gibbs free energy of a $\ce{Br2}(g)$ gas-phase molecule, $c_{\ce{Br-}}$ the ion concentration in mol/l, and $\phi_{\ce{Br}}^{\mathrm{ref}}$ the equilibrium potential for $\ce{Br2}(g)$ evolution at standard conditions. To reference the resulting DFT energies to the experimental Ag/AgCl reference electrode, we shift the values of $\phiE$ using the literature experimental value of $\phi_{\mathrm{E,ref}}^{\mathrm{Ag/AgCl}}=\SI{4.637}{V}$, \cite{jordan1958hydrodynamic} i.e.:
\begin{equation}
    \phiE = \phiE^{\mathrm{vs Ag/AgCl}} + \phi_{\mathrm{E,ref}}^{\mathrm{Ag/AgCl}}.
    \label{eq:ref_AgAgCl}
\end{equation}

Simple substitution of the Gibbs free energy expressions above with DFT energetics ($G\rightarrow E$) ignores vibrational zero-point and temperature effects, which can lead to sizeable errors.\cite{rogal2006ab} To include these efficiently, we reexpress ${g^{\rm \alpha, CHE}_{\rm exc}}$ as\cite{hoermann2021thermodynamic}
\begin{equation}
    \gexc^{\alpha,\mathrm{CHE}} =  \gexc^{\mathrm{clean}} + \theta^{\alpha}_{\ce{Br}} \Gads^{\alpha} \quad ,
    \label{eq:g_exc_to_G_ads}
\end{equation}
where $\gexc^{\mathrm{clean}}$ is the site-normalized cost of creating a clean interface
\begin{equation}
    \gexc^{\mathrm{clean}} = \tfrac{1}{N_{\mathrm{sites}}}[G_{\mathrm{surf},0}^{\mathrm{clean}}- G_{\ce{Ag},0}^{\mathrm{bulk}}] \quad ,
    \label{eq:g_clean_exc}
\end{equation}
and $\Gads^{\alpha}$ is the coverage-normalized adsorption energy for a configuration $\alpha$
\begin{equation}
    \Gads^{\alpha} = \tfrac{1}{N_{\ce{Br}}^{\alpha}} [G_{\mathrm{surf},0}^{\alpha} - G_{\mathrm{surf},0}^{\mathrm{clean}}] - \mu_{\ce{Br}} \quad .
    \label{eq:E_ads}
\end{equation}
As vibrational free energy differences between slabs and equally sized bulk materials largely cancel, we can then approximate $\gexc^{\mathrm{clean}}$ (eq. \ref{eq:g_clean_exc}) with differences in DFT energies. A similar reasoning applies to the energy differences in eq. \ref{eq:E_ads}, which is why it is sufficient to only consider the vibrational modes of ${N^{\alpha}_{\rm Br}}$ adsorbates and of the ${\rm Br_2}$ gas-phase molecule to estimate ${G}^{\alpha}_{\rm ads}$ sufficiently accurate, see SI for details of these vibrational calculations.
\\
As the CHE approximation considers only charge-neutral surfaces (i.e. surfaces at their respective PZC), it intrinsically omits higher-order potential dependencies of the interfacial energetics,\cite{hoermann2019grand, hoermann2021peakpositions} thereby also {\em a priori} fixing the electrosorption valency to its nominal value, $l_{\rm Br} = -1$. These higher-order effects can be approximately included using an efficient IS model, which allows charging the interfacial system. Explicitly evaluating the energetics at applied potential conditions\cite{hoermann2019grand} then yields theoretical predictions for $l_{\rm Br}$\cite{hoermann2021peakpositions}. In fact, the CV current expression of eq. \eqref{eq:CV} emerges naturally from such fully grand-canonical (FGC) energetics\cite{hoermann2021thermodynamic,hoermann2021peakpositions} without adjustable parameters. 
\\
Further analysis of the FGC energetics shows that the higher-order terms are largely captured by adding a second-order, DL charging-related correction term.\cite{hoermann2019grand, hoermann2021thermodynamic} The expression of the free energy within this CHE+DL framework is\cite{hoermann2019grand}
\begin{equation}
    \gexc^{\alpha,\mathrm{CHE+DL}} = \gexc^{\alpha,\mathrm{CHE}} - \frac{1}{2} A_{\mathrm{site}} C^{\alpha}_{0} \left(\phiE - \phi^{\alpha}_{0}\right)^{2} \quad ,
    \label{eq:CHE+DL}
\end{equation}
with $A_{\mathrm{site}}$ the adsorption site-normalized surface area of the substrate, $C^{\alpha}_{0}$ the capacitance, and $\phi^{\alpha}_{0}$ the PZC of structure $\alpha$ (i.e. its work function). 
\\
In contrast to performing explicit calculations at each studied potential, the CHE+DL method allows to capture the dominant potential dependencies in IS environments while necessitating only few, additional DFT evaluations at non-zero surface charge to obtain $C^{\alpha}_{0}$. Please see the SI for a complete listing of all computed work functions, ${G}^{\alpha}_{\rm ads}$ and corresponding $\gexc$.

\subsection{Statistical Sampling}
\label{subsec:statistical_sampling}

Knowing the thermodynamic stability of the set of adsorption structures ${\alpha}$ allows to derive macroscopic observables by an appropriate statistical mechanics treatment that evaluates the configurational entropic contributions. Here, we follow two routes, namely using the previously introduced approach based on mean-field theory (MFT)\cite{hoermann2021thermodynamic} and an approach based on more rigorous lattice GC-MC sampling\cite{koper1998_lattice_gas, koper1998_MC_simulations, Mitchell2001_Static_and_dynamic_MC}. 
\\
Equilibrium coverages within MFT are determined via the construction of an approximate free energy landscape $g^{\rm \theta_{\rm Br}, MFT}$ as a function of $\theta_{\rm Br}$ and subsequent minimization in $\theta_{\rm Br}$-space. In previous work\cite{hoermann2021thermodynamic}, we only considered a single, high-symmetry composition $\alpha$ at each coverage and determined $g^{\rm \theta_{\rm Br}, MFT}$ by interpolating $\gexc^{\theta_{\rm Br},\mathrm{CHE(+DL)}}$ in $\theta_{\rm Br}$ and adding an ideal-solution-like entropy term. Having sampled the full configuration space of the $(4 \times 4)$ supercell, we here construct $g^{\rm \theta_{\ce{Br}}, MFT}$ identically, but instead explicitly average over all configurations $\alpha$ at given $\theta_{\ce{Br}}$ according to
\begin{eqnarray}
\gexc^{\theta_{\ce{Br}},\mathrm{CHE(+DL)}} &=& \sum_{\alpha| \theta^{\alpha}_{\ce{Br}} = \theta_{\ce{Br}}} p^{\alpha} \gexc^{\alpha,\mathrm{CHE(+DL)}} \quad {\rm with} \label{eq:mft_theta_average}
\\
p^{\alpha} &=& \frac{n_{\alpha}}{\sum_{\alpha| \theta^{\alpha}_{\ce{Br}} = \theta_{\ce{Br}} } n_{\alpha} \quad}~~.
\label{eq:multiplicity}
\end{eqnarray}
Here $n_{\alpha}$ is the statistical weight (multiplicity) of each symmetry-inequivalent structure $\alpha$ as determined by enumeration and symmetry reduction of all structures within the $(4 \times 4)$ cell. In the high-temperature and large-cell limit, this explicit average is consistent with the ideal-solution-like entropy term within MFT.\cite{hoermann2019phase}
\\
In our GC-MC calculations, we map the adsorption patterns $\alpha$ on the 2D square lattice of the Ag(100) surface and fit the energetics $\gexc^{\alpha,\mathrm{CHE(+DL)}}$ at given conditions $(\phi_{E},c_{\ce{Br-}})$ with a two-body cluster expansion (2b-CE), using the ICET python package \cite{aangqvist2019icet}, in a similar approach to Ref. \citen{weitzner2017voltage}. Next, we run the GC-MC simulations in a $(18 \times 18)$ 2D square lattice to obtain macroscopic averages for the coverage $\theta_{\ce{Br}}$ at the respective conditions. More computational details and convergence tests are provided in the SI. 
\\
Finally, to derive CV currents from eq. \eqref{eq:CV}, we set $l_{\ce{Br}}=-1$ for the CHE \& MFT and CHE \& GC-MC calculations, and use the analytic expression for $l_{\ce{Br}}$ from Ref. \citen{hoermann2021thermodynamic} for the CHE+DL \& MFT  analysis. For the CHE+DL \& GC-MC analysis, we determine $l_{\ce{Br}}$ via eq. \eqref{eq:electrosorption_valency} with differential coverage changes at points $(\phi_{E}, c_{\ce{Br-}})$  evaluated numerically by performing additional GC-MC simulations at slightly altered conditions $(\phi_{E}+\mathrm{d}\phi_{E}, c_{\ce{Br-}}+\mathrm{d}c_{\ce{Br-}})$, cf. SI for details. 

\section{Results}

In the subsequent sections, we assess the effectiveness of various modeling steps, as illustrated in Fig. \ref{fig:Introduction_alt}. Following the principle of Occam's razor, we start from the most simple and computationally most efficient approach: Vacuum-DFT calculations, CHE thermodynamics, and MFT statistical sampling. By improving the statistical sampling and gradually incorporating solvation and capacitive effects, we carefully examine their influence on the overall outcomes, weighing their potential for improvement against the added complexity and cost they introduce. We always employ the same scan rate and ion concentration as in the normalized experimental CVs of Fig.~\ref{fig:experimental_Br_CVs}, so that the results can be directly benchmarked against this reference.

\subsection{Vacuum Energetics \& CHE: Influence of the Statistical Sampling}

\subsubsection{Robustness of the MFT Approach}

We begin by comparing the performance of MFT and lattice GC-MC. As the influences of the statistical sampling method are largely independent of the \textit{ab initio} thermodynamics modeling and the employed solvation model, we expect the resulting insights to then also transfer to the approaches incorporating solvation and capacitive effects discussed below. Compared to the explicit GC-MC sampling, MFT seems more straightforward and computationally less demanding at first sight. However, as the MFT approach requires representing the coverage-dependent interfacial energetics $\gexc(\theta_{\rm Br})$ as a continuous function, it necessarily involves an interpolation of the discrete first-principles data available at the coverages that can be accessed in the employed finite-size surface unit-cell. Here this is a $(4 \times 4)$ cell which correspondingly provides DFT energetic data at 1/16 ML coverage steps. 

\begin{figure}[htbp]
    \centering
    \includegraphics[width=0.8\textwidth]{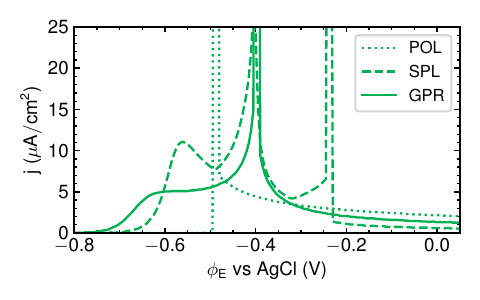}
    \caption{Effect of the employed interpolation method on CVs modeled with vacuum energetics, CHE and MFT. POL = third-order polynomial interpolation, SPL = cubic spline interpolation, and GPR = Gaussian process regression. Only the GPS interpolation recovers the double peak structure observed experimentally, cf. Fig.\ref{fig:experimental_Br_CVs}.}
    \label{fig:MFT_choice_example}
\end{figure}

Due to the strong, repulsive interactions between Br adsorbates, the interpolation method needs to be of a higher order than, for instance, the linear interpolations that have previously been employed for the modeling of CVs of H-electrosorption on Pt\cite{karlberg2007cyclic, mccrum2016ph, mccrum2016first}. To examine the sensitivity of the MFT approach on the employed interpolation method, we therefore compare third-order polynomial (POL), Gaussian process regression (GPR, see SI for more details) and cubic splines (SPL) interpolation. The resulting CVs are shown in Fig. \ref{fig:MFT_choice_example} and are discomfortingly different. While all three methods yield a CV centered around $\sim -0.5$\, V vs AgCl, this CV has a widely differing shape consisting of one, two and three sub-peaks for the POL, GPR and SPL interpolation, respectively.

This finding is easily explained as flexible interpolation methods can lead to non-convex regions in the $\gexc(\theta_{\rm Br})$ function that result in discontinuous coverage changes as a function of the applied potential and thus sharp spikes in the predicted CV.\cite{hoermann2021peakpositions} The POL interpolation introduces no such region, while the more flexible methods GPR and SPL introduce one and two such regions, respectively. We are thus generally faced with the dilemma that a certain flexibility in the interpolation is required to appropriately capture the coverage dependence of $\gexc(\theta_{\rm Br})$, while too much flexibility can quickly lead to artifacts at the given finite DFT data. In principle, this may, of course, be remedied by increasing the $\theta_{\rm Br}$-resolution of the DFT data. Yet, this would involve the use of larger surface unit-cells and more individual calculations, at concomitant strongly increased computational costs. At the present resolution, the GPR is the only method that recovers the experimentally observed double-peak structure of the CV, cf. Fig.~\ref{fig:experimental_Br_CVs}. We ascribe this to the controllable smoothness of the regressive properties of this method, see SI for details, but note that the recovery of the experimental CV shape is only achieved after a careful tuning of the corresponding hyper-parameter. Even though the interpolation step is thus also critical for GPR interpolation, we focus on this method in the following.

\begin{figure}[htbp]
    \centering
    \includegraphics[width=1.\textwidth]{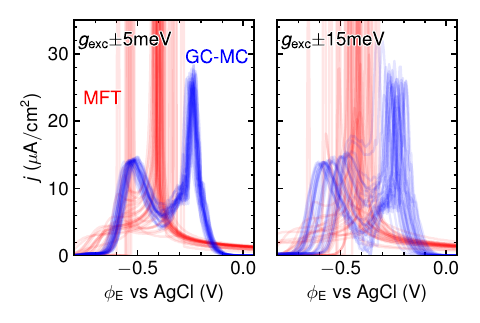}
    \caption{Sensitivity analysis of GPR-interpolated MFT- (red) and GC-MC (blue) CVs (using vacuum energetics and CHE) to white noise in the DFT data. Plotted are 30 CVs each, in which the underlying $\gexc(\theta_{\rm Br})$ DFT data was distorted by random errors in the range (left) $\pm 5$\,meV and (right) $\pm 15$\,meV.}
    \label{fig:Sampling_noise_sensitivity_analysis}
\end{figure}

A second issue for the interpolation method is its robustness to possible noise in the DFT data. Such noise can arise from multiple sources, ranging from not fully converged DFT calculations to finite {\em ab initio} molecular dynamics sampling in explicit solvation models\cite{heenen2020solvation}. Figure~\ref{fig:Sampling_noise_sensitivity_analysis} shows corresponding CVs in which the underlying $G_{\rm ads}^\alpha (\theta_{\rm Br})$ (i.e. the discrete data before interpolation) were distorted with white noise of varying strength (see SI for more noise levels). Again a quite discomforting sensitivity is deduced, in which already small noise levels induce strong shape changes dominated by spikes due to erratic coverage discontinuities. To put this into perspective, we also include in Fig.~\ref{fig:Sampling_noise_sensitivity_analysis} the CVs that are obtained when using the same distorted data as the basis for a GC-MC sampling. Specifically, we here used a 2b-CE with the interaction cutoff set to \SI{4.3}{\angstrom}, such that the expansion includes on-site energy and 1st and 2nd NN interactions that are parametrized with the DFT data, cf. SI for more details on and convergence of the 2b-CE. With the exception of overall CV shifts, the GC-MC CVs retain their peak shape much better under the influence of noise, in fact even up to the high noise level shown in Fig.~\ref{fig:Sampling_noise_sensitivity_analysis}.

The superior stability of GC-MC likely results from the fact that the noise can only affect the interaction weights of the GC-MC's pre-determined Hamiltonian, while it can alter the overall nature of the MFT Hamiltonian. In the present short-range 2b-CE, a change in the adsorption energy only shifts the entire CV peak. The 1st NN interaction is energetically so unfavorable that any small changes do not affect the essential blocking of NN occupations in the adsorbate lattice. As a result, actual variations in the peak shapes are only introduced by noise-induced variations in the weaker repulsive 2nd NN interaction, where increasing or decreasing values merely stretch or compress the CV, cf. SI. This limited mapping induces an inherent robustness to errors. To be fair, one should note though that this is gradually lost when increasing the 2b interaction cutoff or including many-body interactions into the CE. As shown in the SI, we then also obtain somewhat larger distortions to the GC-MC CVs. However, they are never as large as those of the MFT CVs for the same noise level, and we also observe a systematic and rapid convergence of the simulated CVs with respect to an increase in the 2b interaction cutoff. This demonstrates that the robust short-range CE with only 1st and 2nd NN interactions (interaction cutoff set to \SI{4.3}{\angstrom}) as in Fig.~\ref{fig:Sampling_noise_sensitivity_analysis} is fully sufficient for the present system and used henceforth as default.

\subsubsection{MFT vs GC-MC Sampling}

\begin{figure}[htbp]
    \centering
    \includegraphics[width=0.8\textwidth]{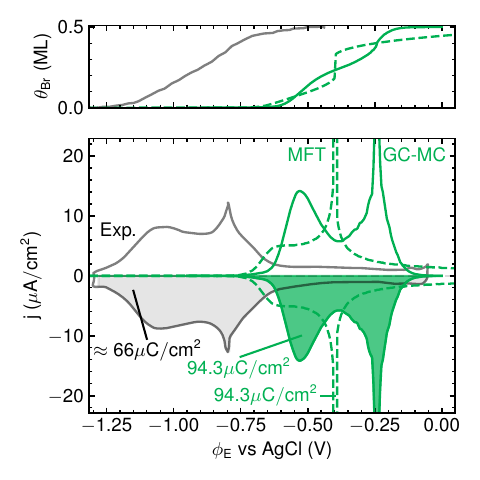}
    \caption{Comparison of the best-practice GPR-interpolated MFT (dashed, green line) and GC-MC (solid green line) CV with the normalized experimental CV from Fig. \ref{fig:experimental_Br_CVs} (solid, gray line)\cite{nakamura2011structure}. Both theoretical CVs are based on vacuum energetics and the CHE. Also indicated is the total transferred electronic charge obtained from integrating each CV. The top panel shows the corresponding surface coverage. The experimental coverage isotherm is taken from chronocoloumetry measurements from Wandlowski \emph{et al.}\cite{wandlowski2001adsorption}.}
    \label{fig:GPR_vs_GCMC}
\end{figure}

The results of the last subsection reveal that while MFT is an easy and quick approach, its sensitivity to the employed interpolation method and to noise in the DFT data render it non-ideal to model CVs with complex peak shapes. This assessment does thereby not even yet extend to its approximate handling of the configurational entropy. We assess the latter in Fig.~\ref{fig:GPR_vs_GCMC} where we directly benchmark the CVs obtained with the determined best-practice MFT and GC-MC model against the normalized experimental data. Both theoretical CVs are strongly shifted and more compressed as compared to the experimental reference. In both methods, the onset of Br electrosorption occurs at $\sim$\SI{-0.6}{\volt} vs AgCl and is followed by a shoulder feature consistent with the experimentally observed peak P1 as discussed in Sec. \ref{sec:experimental_CVs}. Similarly, both methods yield a sharper second peak P2 at higher potentials (MFT at \SI{-0.4}{V~vs~AgCl}, GC-MC at \SI{-0.25}{V~vs~AgCl}). 

In detail, however, the two methods do predict quite different CV shapes, with the MFT approach with its nominally inferior sampling in fact somewhat better reproducing the experimental shape, both in terms of the more hump-like character of the P1 peak and the sharp spike-like character of the P2 peak. Yet, with respect to the latter one can clearly show that this is completely fortuitous. In the GC-MC simulations the P2 peak arises as expected from a second order disorder-order phase transition of the Br adlayer. Using order parameters appropriate for $(2 \times 2)$ ordering\cite{yarnell1973structure,zhang2007accuracy}, the freezing out of the ordered $c(2 \times 2)$ structure from a previously disordered lattice gas at potentials around P2 can nicely be discerned as shown in the SI. In fact, the employed short-range CE truncated to 1st and 2nd NN interactions directly connects to a bulk of work with corresponding model Hamiltonians on square lattices. From such work, the nature of the disorder-order phase transition is well known. For a site-blocking 1st NN repulsive interaction, the transition occurs at about 75-80\% of the limiting coverage of 0.5\,ML.\cite{koper1998_lattice_gas, koper1998_MC_simulations, landau2013statistical, persson1992ordered, taylor1985two} Furthermore, this critical coverage $\theta_{c}$ varies only slightly in the presence of longer-range interactions, and remains at 80\% for a large range of repulsive 2nd NN interaction energies.\cite{taylor1985two} Fully consistent with this, the peak P2 arises at $\theta_{c} \approx 80$\,\% in our GC-MC simulations and the previously discussed robustness of the simulation results in particular with respect to the P2 part of the CV directly correlates with the known robust and universal nature of this phase transition. 

In contrast, MFT is by construction completely agnostic to such disorder-order physics. Here, the P2 peak derives simply from a discontinuous jump in ${\theta_{\ce{Br}}}$ occurring between 0.20 and 0.35\,ML, i.e. at 50-70\,\%  of the maximum coverage. As already stated, this jump is the result of a non-convex coverage-dependence of $\gexc$, and thus depends sensitively on the details of the DFT data points and the interpolation method. The good agreement of the MFT P2 peak shape is thus a prime example of right for the wrong reasons, and we will see next that the worse prediction obtained for the superior GC-MC sampling is in fact the consequence of the hitherto still lacking treatment of solvation and capacitive effects.

\subsection{GC-MC \& CHE: Solvent Stabilization}
\label{subsec:implicit_solvent}

\begin{figure}[htbp]
    \centering
    \includegraphics[width=0.8\textwidth]{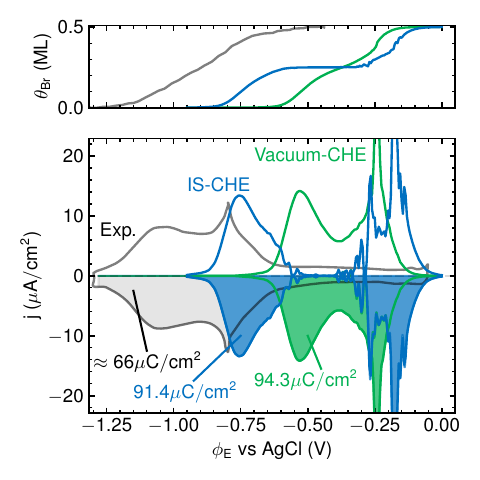}
    \caption{Same as Fig.~\ref{fig:GPR_vs_GCMC}, but now comparing GC-MC \& CHE CVs based on vacuum energetics (solid, green line) and implicit solvation (IS) energetics (solid, blue line) with the normalized experimental CV from Fig. \ref{fig:experimental_Br_CVs} (solid, gray line)\cite{nakamura2011structure}. Also indicated is the total transferred electronic charge obtained from integrating each CV. The top panel shows the corresponding surface coverage. The experimental coverage isotherm is taken from chronocoloumetry measurements from Wandlowski \emph{et al.} \cite{wandlowski2001adsorption}.}
    \label{fig:GCMC_CHE_implicit_solvent}
\end{figure}

In view of the inherent deficiencies of the MFT sampling, we concentrate our ensuing analysis on GC-MC sampling. Apart from the differences in the overall CV shape with respect to the experimental reference, a second discrepancy of the afore discussed GC-MC CV obtained with vacuum energetics and the CHE was an overall offset by $\approx 0.5$\,V. Such a shift to more anodic potentials might well be due to the lack of solvent stabilization in the hitherto employed vacuum energetics. In our next analysis step we correspondingly still stay within the CHE, but now employ the DFT energetics obtained with the implicit solvation model. Figure~\ref{fig:GCMC_CHE_implicit_solvent} compares the corresponding CV with the one obtained with vacuum energetics and the experimental reference. Indeed, the onset of the implicit-solvent CV shifts to lower potentials, reflecting a stabilization of the respective low-coverage adsorbate configurations by the solvent model. However, this is accompanied by an opposite slight upward shift of the higher-coverage P2-peak part of the CV. As a result, the overall CV becomes much broader than the experimental reference and {\em de facto} separates into two parts. 

\begin{figure}[htbp]
    \centering
    \includegraphics[width=0.8\textwidth]{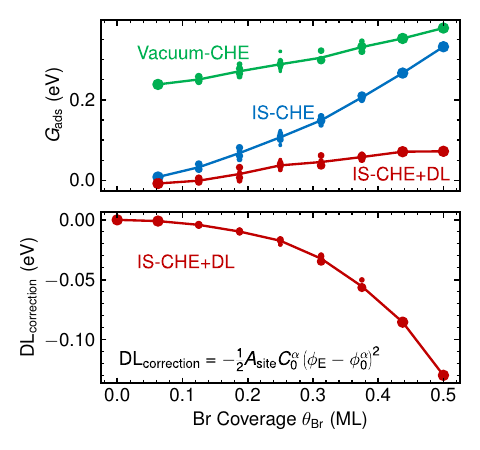}
    \caption{Adsorption energy $\Gads$ for all DFT-calculated configurations $\alpha$ plotted against their respective Br coverage $\theta_{\ce{Br}}$, evaluated at $\phiE = \SI{-0.8}{V~vs~AgCl}$ (top panel), the center of the experimental CV. The size of the scatter points corresponds to $p^{\alpha}$ (eq.~\ref{eq:multiplicity}). We show CHE values for vacuum (green) and implicit solvent energetics (blue), as well as the CHE+DL values within the implicit solvent model (red). The correction term introduced by the CHE+DL scheme (DL$_\mathrm{correction}$) is shown in the bottom panel. In both figures, the lines correspond to the weighted average values at each unique coverage.}
	\label{fig:E_ads_solvent}
\end{figure}

\begin{figure}[htbp]
    \centering
    \includegraphics[width=0.8\textwidth]{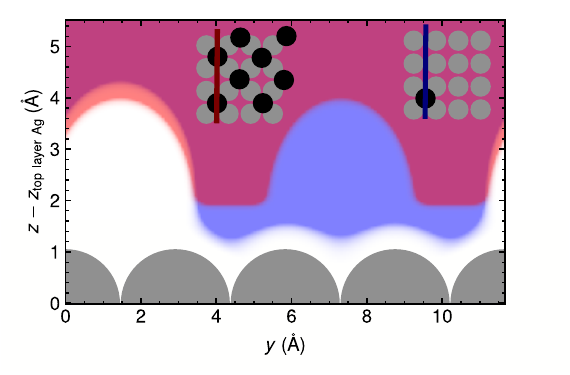}
    \caption{Side view of the solvation cavity of the implicit solvation model for a low-coverage $p(4 \times 4)$ (blue) and a high-coverage $c(2 \times 2)$ (red) Br adsorbate layer. The insets explain the position of the shown vertical cut above the surface. In case of the low-coverage adsorbate layer the dielectric medium extends to much closer to the surface between the adsorbates, thus enabling a higher solvent stabilization due to screening.}
    \label{fig:newfig}
\end{figure}

A direct comparison of the coverage-dependent adsorption energies in vacuum and IS in the top panel in Fig. \ref{fig:E_ads_solvent} points to the origin of this separation. While $\Gads^{\mathrm{CHE}}(\theta_{\rm Br})$ at low coverages are stabilized by the solvent model by $\sim 250$\,meV per Br adsorbate relative to the vacuum energetics, the IS-induced stabilization diminishes with increasing coverage, becoming negligible at the highest coverage of $\theta_{\rm Br} = 0.5$\,ML. %and ultimately even reverses to a slight destabilization at the highest coverage of $\theta_{\rm Br} = 0.5$\,ML.
In the short-range 2b-CE this translates to a decrease in the onsite term of 203\,meV and an doubling of the repulsive 2nd NN interaction term from 60\,meV in vacuum to 135\,meV in implicit solvation. Overall this then spreads the coverage isotherm as seen in Fig.~\ref{fig:GCMC_CHE_implicit_solvent} and concomitantly the CV. The diminishing stabilization in turn is a direct consequence of the implicit solvent representation in form of a dielectric continuum beyond a solvation cavity defined by a threshold electron density\cite{andreussi2012revised}. As apparent from Fig.~\ref{fig:newfig}, at low coverage this cavity extends to close to the surface in the large clean parts of the surface in between the dilute Br adsorbates. In contrast, this is no longer possible at the small spacing between the Br adsorbates at the highest coverage. The stabilization in the IS model results from a simple screening of the repulsive electrostatic interactions between the Br adsorbates by the dielectric medium. With this medium being able to encapsulate the Br adsorbates much better at low coverages, a higher stabilization consequently arises as compared to the high-coverage case where this is no longer possible (as the solvent cannot penetrate between the adsorbates anymore). Even though the IS model is a coarse representation of the true solvation environment, this varying screening and concomitantly differing degrees of solvent stabilization should in principle be the correct physics. As in the case with the sampling before, we thus again arrive at the result that a nominally better modeling does not directly lead to an improved CV observable.

\subsection{GC-MC \& CHE+DL: Capacitive Charging Effects}
\label{sec:DL_effects}
\begin{figure}[htbp]
    \centering
    \includegraphics[width=0.8\textwidth]{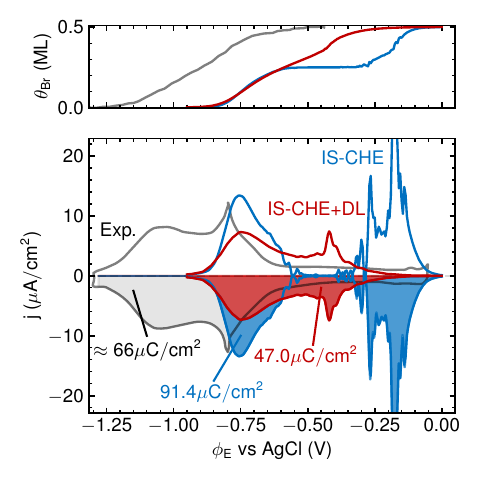}
    \caption{Same as Figs.~\ref{fig:GPR_vs_GCMC} and \ref{fig:GCMC_CHE_implicit_solvent}, but now comparing GC-MC \& implicit solvation CVs based on CHE (solid, blue line) and implicit solvation energetics (solid, red line) with the normalized experimental CV from Fig. \ref{fig:experimental_Br_CVs} (solid, gray line)\cite{nakamura2011structure}. Also indicated is the total transferred electronic charge obtained from integrating each CV. The top panel shows the corresponding surface coverage. The experimental coverage isotherm is taken from chronocoloumetry measurements from Wandlowski \emph{et al.} \cite{wandlowski2001adsorption}.}
    \label{fig:GCMC_CHE+DL}
\end{figure}

The last missing piece in the modeling hierarchy is the consideration of capacitive charging effects via the CHE+DL approach. Figure~\ref{fig:GCMC_CHE+DL} correspondingly compares the simulated GC-MC CV based on implicit solvation energetics at the CHE and CHE+DL level with the experimental reference. Remarkably, the second-order inclusion of the electrode potential largely reverts the excessive CV broadening observed previously when switching from vacuum to implicit solvation energetics at the CHE level, while at the same time leaving the onset potential of the CV unchanged. As a result, a CV shape highly reminiscent of the experimental CV is again obtained, but with the entire CV now also located at more cathodic potentials closer to this reference. 
\\
This result can be rationalized by analyzing the quadratic DL correction term $- \frac{1}{2} A_{\mathrm{site}} C^{\theta_{\rm Br}}_{0} \left(\phiE - \phi^{\theta_{\rm Br}}_{0}\right)^{2}$ that is introduced at this level of theory. Figure~\ref{fig:E_ads_solvent} shows the coverage dependence of this term when approximately evaluating it for $\phiE = -0.8$\,V vs AgCl and thus at a potential that roughly corresponds to the center of the experimental CV. At such relevant potentials, the term becomes increasingly negative with increasing coverage and therefore effectively cancels the increased positive slope of the $\theta_{\rm Br}\Gads^{\rm CHE}$ CHE-term upon changing to implicit solvation energetics, cf. Fig.~\ref{fig:E_ads_solvent}. In consequence and also shown in Fig.~\ref{fig:E_ads_solvent}, $\Gads$ which receives a contribution from both of these terms exhibits almost the same slope with coverage at CHE and vacuum energetics as at CHE+DL and implicit solvation energetics. In other words, the fortuitous agreement of the shape of the CHE plus vacuum energetics CV with experiment was the result of a cancellation of errors introduced by the simultaneously missing solvation and capacitive charging effects. 
\\
However, the CHE+DL approach not only improves the overall peak shape and absolute position of the CV. It also significantly reduces the total transferred electronic charge $\sigma_{\rm Br}$, i.e. the integrated area under the CV, as well as changes the relative height of the P1 and P2 peaks. Both of these changes again improve the comparison to the experimental reference. In particular $\sigma_{\rm Br}$ was consistently overestimated within all previous modeling approaches, cf. Figs.~\ref{fig:GPR_vs_GCMC}, \ref{fig:GCMC_CHE_implicit_solvent} and \ref{fig:GCMC_CHE+DL}, and is now in much better agreement with experiment. Both of these effects arise from the electrosorption valency $l_{\rm Br}\left(\theta_{\mathrm{Br}}, \phiE\right)$ that scales the overall CV, cf. eq.~\ref{eq:CV}, and that in the CHE+DL approach can now take values less negative than the nominal charge of -1.\cite{hoermann2019grand} As shown in the SI, the CHE+DL $l_{\rm Br}$ is in fact not constant, but increases almost linearly from -0.7 to -0.45 over the potential window (aka coverage) of the CV and falls thus into the range estimated for the electrosorption valency from the experimental data, cf. Sec. \ref{sec:experimental_CVs}. This potential dependence of $l_{\rm Br}$ then alters the relative heights of the P1 and P2 peaks, as less charge is transmitted per adsorbate at lower than at higher coverages. It is also only this non-integer value of $l_{\rm Br}$ that leads to the non-Nernstian potential shift of the P2 peak with \ce{Br-} concentration reported experimentally\cite{wandlowski2001adsorption}.
\\
Overall and gratifyingly, it is thus indeed the CV modelled at the nominally best level of theory that achieves the best agreement with the experimental reference, i.e. a CV obtained by GC-MC sampling, an energetics accounting for solvation effects at least at the level of an implicit solvation model, as well as considering capacitive charging effects to second order. In fact, considering that we have focused only on computationally efficient approaches that in many respects are still effective -- prominently the description of the solvation environment by a mere dielectric continuum -- this agreement down to width, shape, and integrated area of the CV is quite impressive. What remains as the largest discrepancy is the overall potential shift of about $\sim 0.3$\,V of the predicted CV vs the experimental data. We ascribe much of this difference to the employed semi-local PBE DFT functional and support this assignment with a recalculation of all vacuum DFT energetics with the revPBE functional, cf. SI for details. We obtain $\Gads$ for all configurations $\alpha$ that are predominantly shifted by about $\sim +0.15$\,eV as compared to the corresponding PBE values. In consequence, a short-range 2b-CE based on this energetics exhibits largely unchanged 1st and 2nd NN interactions, but instead only an onsite term that is less stable by $\sim 0.15 $\,eV. Obviously, the entire analysis of the last sections would thus hold in an analogous way for this CE, just with the entire simulated CVs shifted by $\sim 0.15$\,V to more cathodic potentials and thus even further away from the experimental reference. This agrees with the general expectation of an even weaker binding at the revPBE level and the knowledge that already the PBE underestimates the binding of halides.\cite{friedrich2019coordination, wang2021framework, schmidt2018benchmark, wellendorff2015benchmark} Of course, just testing one other semi-local functional does not do justice to the wealth of approximate DFT energetics that can in principle be obtained. Nevertheless, we believe that the provided singular example illustrates that this uncertainty in the energetics may prominently lead to overall shifts of the simulated CV. As such, the approximate DFT energetics is in our view the most likely candidate to explain the remaining discrepancy of the GC-MC CHE+DL CV based on implicit solvation energetics with respect to the experimental reference. 

\section{Summary and Conclusions}

In this benchmark study we have systematically analyzed prominent choices in the simulation workflow for thermodynamic CVs, using Br electrosorption at a model Ag(100) electrode as a representative showcase. Focusing on computationally efficient, prevalent approaches, we analyzed the influence of an approximate account of the solvation environment in form of energetics calculated within an implicit solvation model, of an {\em ab initio} thermodynamics description that incorporates capacitive charging up to second order in the potential, as well as of a grand-canonical Monte Carlo sampling that explicitly evaluates configurational entropic effects in the adlayer. As a crucial insight, we observed an intricate error cancellation when several of these aspects were treated more approximately. A good agreement of a simulated CV with experimental data can thus not uncritically be taken as evidence that the employed level of theory was sufficient.

At the nominally best level of theory considered in this study (GC-MC sampling, implicit solvation energetics and CHE+DL thermodynamics) we obtain a gratifying essentially quantitative agreement of the simulated CV with experimental reference data. The analysis provided suggests that this is the result of an appropriate description of key physics of this system, in particular a coverage-dependent solvation stabilization due to a varying capability of the solvent to penetrate the adlayer and the disorder-order phase transition of the Br adlayer at higher coverages. Nevertheless, in view of the error cancellations observed at the lower levels of theory, this agreement should be scrutinized further in future work. Most prominently, we envision explicit electrolyte approaches as the next frontier that would provide most valuable feedback on the true reliability of the here employed implicit solvation method. Specifically, we hereby refer to both the parametrization of the implicit solvation model, as well as its fundamental deficiencies in appropriately describing H-bonding networks and other directed solvent interactions at all. We consider the wealth of experimental CVs available for this system as an opportunity to systematically analyze such aspects with respect to a firm reference. 

\begin{acknowledgement}
The authors thank S. Beinlich and T. Eggert for useful discussions and suggestions during this project, as well as V. J. Bukas and H. Oschinski for their contributions to the development of this manuscript. The authors acknowledge funding and support from the German Research Foundation (DFG) under Germany's Excellence Strategy - EXC 2089/1- 390776260 (e-conversion)
%%% From Nico's JCTC Thermodynamic CV paper
and financial support through the EuroTech Postdoc Programme, which is co-funded by the European Commission under its framework programme Horizon 2020 and Grant Agreement number 754462.
All computations were performed on the HPC system Raven at the Max Planck Computing and Data Facility, which we gratefully acknowledge.
\end{acknowledgement}

\bibliography{ref.bib}

\providecommand{\latin}[1]{#1}
\makeatletter
\providecommand{\doi}
  {\begingroup\let\do\@makeother\dospecials
  \catcode`\{=1 \catcode`\}=2 \doi@aux}
\providecommand{\doi@aux}[1]{\endgroup\texttt{#1}}
\makeatother
\providecommand*\mcitethebibliography{\thebibliography}
\csname @ifundefined\endcsname{endmcitethebibliography}
  {\let\endmcitethebibliography\endthebibliography}{}
\begin{mcitethebibliography}{62}
\providecommand*\natexlab[1]{#1}
\providecommand*\mciteSetBstSublistMode[1]{}
\providecommand*\mciteSetBstMaxWidthForm[2]{}
\providecommand*\mciteBstWouldAddEndPuncttrue
  {\def\EndOfBibitem{\unskip.}}
\providecommand*\mciteBstWouldAddEndPunctfalse
  {\let\EndOfBibitem\relax}
\providecommand*\mciteSetBstMidEndSepPunct[3]{}
\providecommand*\mciteSetBstSublistLabelBeginEnd[3]{}
\providecommand*\EndOfBibitem{}
\mciteSetBstSublistMode{f}
\mciteSetBstMaxWidthForm{subitem}{(\alph{mcitesubitemcount})}
\mciteSetBstSublistLabelBeginEnd
  {\mcitemaxwidthsubitemform\space}
  {\relax}
  {\relax}

\bibitem[Bard and Zoski(2000)Bard, and Zoski]{bard2000voltammetry}
Bard,~A.~J.; Zoski,~C.~G. Voltammetry Retrospective. \emph{Analytical
  Chemistry} \textbf{2000}, \emph{72}, 346 A--352 A\relax
\mciteBstWouldAddEndPuncttrue
\mciteSetBstMidEndSepPunct{\mcitedefaultmidpunct}
{\mcitedefaultendpunct}{\mcitedefaultseppunct}\relax
\EndOfBibitem
\bibitem[Nicholson(1965)]{nicholson1965theory}
Nicholson,~R.~S. Theory and Application of Cyclic Voltammetry for Measurement
  of Electrode Reaction Kinetics. \emph{Analytical Chemistry} \textbf{1965},
  \emph{37}, 1351--1355\relax
\mciteBstWouldAddEndPuncttrue
\mciteSetBstMidEndSepPunct{\mcitedefaultmidpunct}
{\mcitedefaultendpunct}{\mcitedefaultseppunct}\relax
\EndOfBibitem
\bibitem[Elgrishi \latin{et~al.}(2018)Elgrishi, Rountree, McCarthy, Rountree,
  Eisenhart, and Dempsey]{elgrishi2018practical}
Elgrishi,~N.; Rountree,~K.~J.; McCarthy,~B.~D.; Rountree,~E.~S.;
  Eisenhart,~T.~T.; Dempsey,~J.~L. A Practical Beginner’s Guide to Cyclic
  Voltammetry. \emph{Journal of Chemical Education} \textbf{2018}, \emph{95},
  197--206\relax
\mciteBstWouldAddEndPuncttrue
\mciteSetBstMidEndSepPunct{\mcitedefaultmidpunct}
{\mcitedefaultendpunct}{\mcitedefaultseppunct}\relax
\EndOfBibitem
\bibitem[Kissinger and Heineman(1983)Kissinger, and
  Heineman]{kissinger1983cyclic}
Kissinger,~P.~T.; Heineman,~W.~R. Cyclic voltammetry. \emph{Journal of Chemical
  Education} \textbf{1983}, \emph{60}, 702\relax
\mciteBstWouldAddEndPuncttrue
\mciteSetBstMidEndSepPunct{\mcitedefaultmidpunct}
{\mcitedefaultendpunct}{\mcitedefaultseppunct}\relax
\EndOfBibitem
\bibitem[Climent and Feliu(2018)Climent, and Feliu]{climent2018cyclic}
Climent,~V.; Feliu,~J. In \emph{Encyclopedia of Interfacial Chemistry};
  Wandelt,~K., Ed.; Elsevier: Oxford, 2018; pp 48--74\relax
\mciteBstWouldAddEndPuncttrue
\mciteSetBstMidEndSepPunct{\mcitedefaultmidpunct}
{\mcitedefaultendpunct}{\mcitedefaultseppunct}\relax
\EndOfBibitem
\bibitem[Engstfeld \latin{et~al.}(2018)Engstfeld, Maagaard, Horch,
  Chorkendorff, and Stephens]{engstfeld2018polycrystalline}
Engstfeld,~A.~K.; Maagaard,~T.; Horch,~S.; Chorkendorff,~I.; Stephens,~I. E.~L.
  Polycrystalline and Single-Crystal Cu Electrodes: Influence of Experimental
  Conditions on the Electrochemical Properties in Alkaline Media.
  \emph{Chemistry – A European Journal} \textbf{2018}, \emph{24},
  17743--17755\relax
\mciteBstWouldAddEndPuncttrue
\mciteSetBstMidEndSepPunct{\mcitedefaultmidpunct}
{\mcitedefaultendpunct}{\mcitedefaultseppunct}\relax
\EndOfBibitem
\bibitem[Sheng \latin{et~al.}(2015)Sheng, Zhuang, Gao, Zheng, Chen, and
  Yan]{sheng2015correlating}
Sheng,~W.; Zhuang,~Z.; Gao,~M.; Zheng,~J.; Chen,~J.~G.; Yan,~Y. Correlating
  hydrogen oxidation and evolution activity on platinum at different pH with
  measured hydrogen binding energy. \emph{Nature Communications} \textbf{2015},
  \emph{6}, 5848\relax
\mciteBstWouldAddEndPuncttrue
\mciteSetBstMidEndSepPunct{\mcitedefaultmidpunct}
{\mcitedefaultendpunct}{\mcitedefaultseppunct}\relax
\EndOfBibitem
\bibitem[Aristov and Habekost(2015)Aristov, and Habekost]{aristov2015cyclic}
Aristov,~N.; Habekost,~A. Cyclic Voltammetry - A Versatile Electrochemical
  Method Investigating Electron Transfer Processes. \emph{World Journal of
  Chemical Education} \textbf{2015}, \emph{3}, 115--119\relax
\mciteBstWouldAddEndPuncttrue
\mciteSetBstMidEndSepPunct{\mcitedefaultmidpunct}
{\mcitedefaultendpunct}{\mcitedefaultseppunct}\relax
\EndOfBibitem
\bibitem[Karlberg \latin{et~al.}(2007)Karlberg, Jaramillo, Sk\'ulason,
  Rossmeisl, Bligaard, and N{\o}rskov]{karlberg2007cyclic}
Karlberg,~G.~S.; Jaramillo,~T.~F.; Sk\'ulason,~E.; Rossmeisl,~J.; Bligaard,~T.;
  N{\o}rskov,~J.~K. Cyclic Voltammograms for H on Pt(111) and Pt(100) from
  First Principles. \emph{Physical Review Letters} \textbf{2007}, \emph{99},
  126101\relax
\mciteBstWouldAddEndPuncttrue
\mciteSetBstMidEndSepPunct{\mcitedefaultmidpunct}
{\mcitedefaultendpunct}{\mcitedefaultseppunct}\relax
\EndOfBibitem
\bibitem[H\"{o}rmann and Reuter(2021)H\"{o}rmann, and
  Reuter]{hoermann2021thermodynamic}
H\"{o}rmann,~N.~G.; Reuter,~K. Thermodynamic Cyclic Voltammograms Based on
  \textit{Ab Initio} Calculations: Ag(111) in Halide-Containing Solutions.
  \emph{Journal of Chemical Theory and Computation} \textbf{2021}, \emph{17},
  1782--1794, PMID: 33606513\relax
\mciteBstWouldAddEndPuncttrue
\mciteSetBstMidEndSepPunct{\mcitedefaultmidpunct}
{\mcitedefaultendpunct}{\mcitedefaultseppunct}\relax
\EndOfBibitem
\bibitem[Schultze and Vetter(1973)Schultze, and
  Vetter]{schultze1973experimental}
Schultze,~J.; Vetter,~K. Experimental determination and interpretation of the
  electrosorption valency $\gamma$. \emph{Journal of Electroanalytical
  Chemistry and Interfacial Electrochemistry} \textbf{1973}, \emph{44},
  63--81\relax
\mciteBstWouldAddEndPuncttrue
\mciteSetBstMidEndSepPunct{\mcitedefaultmidpunct}
{\mcitedefaultendpunct}{\mcitedefaultseppunct}\relax
\EndOfBibitem
\bibitem[Guidelli and Schmickler(2005)Guidelli, and
  Schmickler]{guidelli2005electrosorption}
Guidelli,~R.; Schmickler,~W. In \emph{Modern Aspects of Electrochemistry};
  Conway,~B.~E., Vayenas,~C.~G., White,~R.~E., Gamboa-Adelco,~M.~E., Eds.;
  Springer US: Boston, MA, 2005; pp 303--371\relax
\mciteBstWouldAddEndPuncttrue
\mciteSetBstMidEndSepPunct{\mcitedefaultmidpunct}
{\mcitedefaultendpunct}{\mcitedefaultseppunct}\relax
\EndOfBibitem
\bibitem[Wandlowski \latin{et~al.}(2001)Wandlowski, Wang, and
  Ocko]{wandlowski2001adsorption}
Wandlowski,~T.; Wang,~J.; Ocko,~B. Adsorption of bromide at the Ag(100)
  electrode surface. \emph{Journal of Electroanalytical Chemistry}
  \textbf{2001}, \emph{500}, 418--434\relax
\mciteBstWouldAddEndPuncttrue
\mciteSetBstMidEndSepPunct{\mcitedefaultmidpunct}
{\mcitedefaultendpunct}{\mcitedefaultseppunct}\relax
\EndOfBibitem
\bibitem[Koper(1998)]{koper1998_lattice_gas}
Koper,~M.~T. A lattice-gas model for halide adsorption on single-crystal
  electrodes. \emph{Journal of Electroanalytical Chemistry} \textbf{1998},
  \emph{450}, 189--201\relax
\mciteBstWouldAddEndPuncttrue
\mciteSetBstMidEndSepPunct{\mcitedefaultmidpunct}
{\mcitedefaultendpunct}{\mcitedefaultseppunct}\relax
\EndOfBibitem
\bibitem[Nakamura \latin{et~al.}(2011)Nakamura, Nakajima, Sato, Hoshi, and
  Sakata]{nakamura2011structure}
Nakamura,~M.; Nakajima,~Y.; Sato,~N.; Hoshi,~N.; Sakata,~O. Structure of the
  electrical double layer on Ag(100): Promotive effect of cationic species on
  Br adlayer formation. \emph{Physical Review B} \textbf{2011}, \emph{84},
  165433\relax
\mciteBstWouldAddEndPuncttrue
\mciteSetBstMidEndSepPunct{\mcitedefaultmidpunct}
{\mcitedefaultendpunct}{\mcitedefaultseppunct}\relax
\EndOfBibitem
\bibitem[N{\o}rskov \latin{et~al.}(2004)N{\o}rskov, Rossmeisl, Logadottir,
  Lindqvist, Kitchin, Bligaard, and J\'{o}nsson]{norskov2004origin}
N{\o}rskov,~J.~K.; Rossmeisl,~J.; Logadottir,~A.; Lindqvist,~L.;
  Kitchin,~J.~R.; Bligaard,~T.; J\'{o}nsson,~H. Origin of the Overpotential for
  Oxygen Reduction at a Fuel-Cell Cathode. \emph{The Journal of Physical
  Chemistry B} \textbf{2004}, \emph{108}, 17886--17892\relax
\mciteBstWouldAddEndPuncttrue
\mciteSetBstMidEndSepPunct{\mcitedefaultmidpunct}
{\mcitedefaultendpunct}{\mcitedefaultseppunct}\relax
\EndOfBibitem
\bibitem[H{\"o}rmann \latin{et~al.}(2020)H{\"o}rmann, Marzari, and
  Reuter]{hormann2020electrosorption}
H{\"o}rmann,~N.~G.; Marzari,~N.; Reuter,~K. Electrosorption at metal surfaces
  from first principles. \emph{npj Computational Materials} \textbf{2020},
  \emph{6}, 136\relax
\mciteBstWouldAddEndPuncttrue
\mciteSetBstMidEndSepPunct{\mcitedefaultmidpunct}
{\mcitedefaultendpunct}{\mcitedefaultseppunct}\relax
\EndOfBibitem
\bibitem[Engstfeld \latin{et~al.}(2023)Engstfeld, Rüth, linuxrider, and
  H\"{o}rmann]{engstfeld2023echemdb}
Engstfeld,~A.; Rüth,~J.; linuxrider,; H\"{o}rmann,~N.~G. echemdb/echemdb:
  0.6.0. 2023; \url{https://doi.org/10.5281/zenodo.7834993}\relax
\mciteBstWouldAddEndPuncttrue
\mciteSetBstMidEndSepPunct{\mcitedefaultmidpunct}
{\mcitedefaultendpunct}{\mcitedefaultseppunct}\relax
\EndOfBibitem
\bibitem[Endo \latin{et~al.}(1999)Endo, Kiguchi, Yokoyama, Ito, and
  Ohta]{endo1999situ}
Endo,~O.; Kiguchi,~M.; Yokoyama,~T.; Ito,~M.; Ohta,~T. In-situ X-ray absorption
  studies of bromine on the Ag(100) electrode. \emph{Journal of
  Electroanalytical Chemistry} \textbf{1999}, \emph{473}, 19--24\relax
\mciteBstWouldAddEndPuncttrue
\mciteSetBstMidEndSepPunct{\mcitedefaultmidpunct}
{\mcitedefaultendpunct}{\mcitedefaultseppunct}\relax
\EndOfBibitem
\bibitem[Ocko \latin{et~al.}(1997)Ocko, Wang, and Wandlowski]{ocko1997bromide}
Ocko,~B.~M.; Wang,~J.~X.; Wandlowski,~T. Bromide Adsorption on Ag(001): A
  Potential Induced Two-Dimensional Ising Order-Disorder Transition.
  \emph{Physical Review Letters} \textbf{1997}, \emph{79}, 1511--1514\relax
\mciteBstWouldAddEndPuncttrue
\mciteSetBstMidEndSepPunct{\mcitedefaultmidpunct}
{\mcitedefaultendpunct}{\mcitedefaultseppunct}\relax
\EndOfBibitem
\bibitem[Koper(1998)]{koper1998_MC_simulations}
Koper,~M.~T. Monte Carlo simulations of ionic adsorption isotherms at
  single-crystal electrodes. \emph{Electrochimica Acta} \textbf{1998},
  \emph{44}, 1207--1212\relax
\mciteBstWouldAddEndPuncttrue
\mciteSetBstMidEndSepPunct{\mcitedefaultmidpunct}
{\mcitedefaultendpunct}{\mcitedefaultseppunct}\relax
\EndOfBibitem
\bibitem[Wang and Rikvold(2002)Wang, and Rikvold]{wang2002ab}
Wang,~S.; Rikvold,~P.~A. Ab initio calculations for bromine adlayers on the
  Ag(100) and Au(100) surfaces: The $c(2\ifmmode\times\else\texttimes\fi{}2)$
  structure. \emph{Physical Review B} \textbf{2002}, \emph{65}, 155406\relax
\mciteBstWouldAddEndPuncttrue
\mciteSetBstMidEndSepPunct{\mcitedefaultmidpunct}
{\mcitedefaultendpunct}{\mcitedefaultseppunct}\relax
\EndOfBibitem
\bibitem[Mitchell \latin{et~al.}(2000)Mitchell, Brown, and
  Rikvold]{mitchell2000dynamics}
Mitchell,~S.; Brown,~G.; Rikvold,~P. Dynamics of Br electrosorption on
  single-crystal Ag(100): a computational study. \emph{Journal of
  Electroanalytical Chemistry} \textbf{2000}, \emph{493}, 68--74\relax
\mciteBstWouldAddEndPuncttrue
\mciteSetBstMidEndSepPunct{\mcitedefaultmidpunct}
{\mcitedefaultendpunct}{\mcitedefaultseppunct}\relax
\EndOfBibitem
\bibitem[Mitchell \latin{et~al.}(2002)Mitchell, Wang, and
  Rikvold]{mitchell2002halide}
Mitchell,~S.~J.; Wang,~S.; Rikvold,~P.~A. Halide adsorption on single-crystal
  silver substrates: dynamic simulations and ab initio density functional
  theory. \emph{Faraday Discussions} \textbf{2002}, \emph{121}, 53--69\relax
\mciteBstWouldAddEndPuncttrue
\mciteSetBstMidEndSepPunct{\mcitedefaultmidpunct}
{\mcitedefaultendpunct}{\mcitedefaultseppunct}\relax
\EndOfBibitem
\bibitem[Persson(1992)]{persson1992ordered}
Persson,~B. Ordered structures and phase transitions in adsorbed layers.
  \emph{Surface Science Reports} \textbf{1992}, \emph{15}, 1--135\relax
\mciteBstWouldAddEndPuncttrue
\mciteSetBstMidEndSepPunct{\mcitedefaultmidpunct}
{\mcitedefaultendpunct}{\mcitedefaultseppunct}\relax
\EndOfBibitem
\bibitem[Landau and Lifshitz(1980)Landau, and Lifshitz]{landau2013statistical}
Landau,~L.~D.; Lifshitz,~E.~M. In \emph{Statistical Physics}, third edition
  ed.; Landau,~L.~D., Lifshitz,~E.~M., Eds.; Butterworth-Heinemann: Oxford,
  1980; pp 446--516\relax
\mciteBstWouldAddEndPuncttrue
\mciteSetBstMidEndSepPunct{\mcitedefaultmidpunct}
{\mcitedefaultendpunct}{\mcitedefaultseppunct}\relax
\EndOfBibitem
\bibitem[H{\"o}rmann and Reuter(2021)H{\"o}rmann, and
  Reuter]{hoermann2021peakpositions}
H{\"o}rmann,~N.~G.; Reuter,~K. Thermodynamic cyclic voltammograms: peak
  positions and shapes. \emph{Journal of Physics: Condensed Matter}
  \textbf{2021}, \emph{33}, 264004\relax
\mciteBstWouldAddEndPuncttrue
\mciteSetBstMidEndSepPunct{\mcitedefaultmidpunct}
{\mcitedefaultendpunct}{\mcitedefaultseppunct}\relax
\EndOfBibitem
\bibitem[Koper and Lukkien(2000)Koper, and Lukkien]{koper2000modeling}
Koper,~M.~T.; Lukkien,~J.~J. Modeling the butterfly: the voltammetry of
  ($\sqrt{3}\times\sqrt{3}$)R30$^{\circ}$ and p($2\times2$) overlayers on (111)
  electrodes. \emph{Journal of Electroanalytical Chemistry} \textbf{2000},
  \emph{485}, 161--165\relax
\mciteBstWouldAddEndPuncttrue
\mciteSetBstMidEndSepPunct{\mcitedefaultmidpunct}
{\mcitedefaultendpunct}{\mcitedefaultseppunct}\relax
\EndOfBibitem
\bibitem[Perdew \latin{et~al.}(1996)Perdew, Burke, and
  Ernzerhof]{perdew1996generalized}
Perdew,~J.~P.; Burke,~K.; Ernzerhof,~M. Generalized Gradient Approximation Made
  Simple. \emph{Physical Review Letters} \textbf{1996}, \emph{77},
  3865--3868\relax
\mciteBstWouldAddEndPuncttrue
\mciteSetBstMidEndSepPunct{\mcitedefaultmidpunct}
{\mcitedefaultendpunct}{\mcitedefaultseppunct}\relax
\EndOfBibitem
\bibitem[Giannozzi \latin{et~al.}(2009)Giannozzi, Baroni, Bonini, Calandra,
  Car, Cavazzoni, Ceresoli, Chiarotti, Cococcioni, Dabo, Corso, de~Gironcoli,
  Fabris, Fratesi, Gebauer, Gerstmann, Gougoussis, Kokalj, Lazzeri,
  Martin-Samos, Marzari, Mauri, Mazzarello, Paolini, Pasquarello, Paulatto,
  Sbraccia, Scandolo, Sclauzero, Seitsonen, Smogunov, Umari, and
  Wentzcovitch]{giannozzi2009p}
Giannozzi,~P. \latin{et~al.}  QUANTUM ESPRESSO: a modular and open-source
  software project for quantum simulations of materials. \emph{Journal of
  Physics: Condensed Matter} \textbf{2009}, \emph{21}, 395502\relax
\mciteBstWouldAddEndPuncttrue
\mciteSetBstMidEndSepPunct{\mcitedefaultmidpunct}
{\mcitedefaultendpunct}{\mcitedefaultseppunct}\relax
\EndOfBibitem
\bibitem[Giannozzi \latin{et~al.}(2017)Giannozzi, Andreussi, Brumme, Bunau,
  Nardelli, Calandra, Car, Cavazzoni, Ceresoli, Cococcioni, Colonna, Carnimeo,
  Corso, de~Gironcoli, Delugas, DiStasio, Ferretti, Floris, Fratesi, Fugallo,
  Gebauer, Gerstmann, Giustino, Gorni, Jia, Kawamura, Ko, Kokalj,
  Küçükbenli, Lazzeri, Marsili, Marzari, Mauri, Nguyen, Nguyen, de-la Roza,
  Paulatto, Poncé, Rocca, Sabatini, Santra, Schlipf, Seitsonen, Smogunov,
  Timrov, Thonhauser, Umari, Vast, Wu, and Baroni]{giannozzi2017advanced}
Giannozzi,~P. \latin{et~al.}  Advanced capabilities for materials modelling
  with Quantum ESPRESSO. \emph{Journal of Physics: Condensed Matter}
  \textbf{2017}, \emph{29}, 465901\relax
\mciteBstWouldAddEndPuncttrue
\mciteSetBstMidEndSepPunct{\mcitedefaultmidpunct}
{\mcitedefaultendpunct}{\mcitedefaultseppunct}\relax
\EndOfBibitem
\bibitem[Garrity \latin{et~al.}(2014)Garrity, Bennett, Rabe, and
  Vanderbilt]{garrity2014pseudopotentials}
Garrity,~K.~F.; Bennett,~J.~W.; Rabe,~K.~M.; Vanderbilt,~D. Pseudopotentials
  for high-throughput DFT calculations. \emph{Computational Materials Science}
  \textbf{2014}, \emph{81}, 446--452\relax
\mciteBstWouldAddEndPuncttrue
\mciteSetBstMidEndSepPunct{\mcitedefaultmidpunct}
{\mcitedefaultendpunct}{\mcitedefaultseppunct}\relax
\EndOfBibitem
\bibitem[Huber(2022)]{huber2022automated}
Huber,~S.~P. Automated reproducible workflows and data provenance with AiiDA.
  \emph{Nature Reviews Physics} \textbf{2022}, \emph{4}, 431--431\relax
\mciteBstWouldAddEndPuncttrue
\mciteSetBstMidEndSepPunct{\mcitedefaultmidpunct}
{\mcitedefaultendpunct}{\mcitedefaultseppunct}\relax
\EndOfBibitem
\bibitem[Andreussi \latin{et~al.}(2012)Andreussi, Dabo, and
  Marzari]{andreussi2012revised}
Andreussi,~O.; Dabo,~I.; Marzari,~N. Revised self-consistent continuum
  solvation in electronic-structure calculations. \emph{The Journal of Chemical
  Physics} \textbf{2012}, \emph{136}, 064102\relax
\mciteBstWouldAddEndPuncttrue
\mciteSetBstMidEndSepPunct{\mcitedefaultmidpunct}
{\mcitedefaultendpunct}{\mcitedefaultseppunct}\relax
\EndOfBibitem
\bibitem[H\"{o}rmann \latin{et~al.}(2019)H\"{o}rmann, Andreussi, and
  Marzari]{hoermann2019grand}
H\"{o}rmann,~N.~G.; Andreussi,~O.; Marzari,~N. Grand canonical simulations of
  electrochemical interfaces in implicit solvation models. \emph{The Journal of
  Chemical Physics} \textbf{2019}, \emph{150}, 041730\relax
\mciteBstWouldAddEndPuncttrue
\mciteSetBstMidEndSepPunct{\mcitedefaultmidpunct}
{\mcitedefaultendpunct}{\mcitedefaultseppunct}\relax
\EndOfBibitem
\bibitem[Schimka \latin{et~al.}(2010)Schimka, Harl, Stroppa, Gr{\"u}neis,
  Marsman, Mittendorfer, and Kresse]{schimka2010accurate}
Schimka,~L.; Harl,~J.; Stroppa,~A.; Gr{\"u}neis,~A.; Marsman,~M.;
  Mittendorfer,~F.; Kresse,~G. Accurate surface and adsorption energies from
  many-body perturbation theory. \emph{Nature Materials} \textbf{2010},
  \emph{9}, 741--744\relax
\mciteBstWouldAddEndPuncttrue
\mciteSetBstMidEndSepPunct{\mcitedefaultmidpunct}
{\mcitedefaultendpunct}{\mcitedefaultseppunct}\relax
\EndOfBibitem
\bibitem[Schmidt and Thygesen(2018)Schmidt, and Thygesen]{schmidt2018benchmark}
Schmidt,~P.~S.; Thygesen,~K.~S. Benchmark Database of Transition Metal Surface
  and Adsorption Energies from Many-Body Perturbation Theory. \emph{The Journal
  of Physical Chemistry C} \textbf{2018}, \emph{122}, 4381--4390\relax
\mciteBstWouldAddEndPuncttrue
\mciteSetBstMidEndSepPunct{\mcitedefaultmidpunct}
{\mcitedefaultendpunct}{\mcitedefaultseppunct}\relax
\EndOfBibitem
\bibitem[Friedrich \latin{et~al.}(2019)Friedrich, Usanmaz, Oses, Supka,
  Fornari, Buongiorno~Nardelli, Toher, and
  Curtarolo]{friedrich2019coordination}
Friedrich,~R.; Usanmaz,~D.; Oses,~C.; Supka,~A.; Fornari,~M.;
  Buongiorno~Nardelli,~M.; Toher,~C.; Curtarolo,~S. Coordination corrected ab
  initio formation enthalpies. \emph{npj Computational Materials}
  \textbf{2019}, \emph{5}, 1--12\relax
\mciteBstWouldAddEndPuncttrue
\mciteSetBstMidEndSepPunct{\mcitedefaultmidpunct}
{\mcitedefaultendpunct}{\mcitedefaultseppunct}\relax
\EndOfBibitem
\bibitem[Wang \latin{et~al.}(2021)Wang, Kingsbury, McDermott, Horton, Jain,
  Ong, Dwaraknath, and Persson]{wang2021framework}
Wang,~A.; Kingsbury,~R.; McDermott,~M.; Horton,~M.; Jain,~A.; Ong,~S.~P.;
  Dwaraknath,~S.; Persson,~K.~A. A framework for quantifying uncertainty in DFT
  energy corrections. \emph{Scientific Reports} \textbf{2021}, \emph{11},
  1--10\relax
\mciteBstWouldAddEndPuncttrue
\mciteSetBstMidEndSepPunct{\mcitedefaultmidpunct}
{\mcitedefaultendpunct}{\mcitedefaultseppunct}\relax
\EndOfBibitem
\bibitem[Wellendorff \latin{et~al.}(2015)Wellendorff, Silbaugh, Garcia-Pintos,
  N{\o}rskov, Bligaard, Studt, and Campbell]{wellendorff2015benchmark}
Wellendorff,~J.; Silbaugh,~T.~L.; Garcia-Pintos,~D.; N{\o}rskov,~J.~K.;
  Bligaard,~T.; Studt,~F.; Campbell,~C.~T. A benchmark database for adsorption
  bond energies to transition metal surfaces and comparison to selected DFT
  functionals. \emph{Surface Science} \textbf{2015}, \emph{640}, 36--44,
  Reactivity Concepts at Surfaces: Coupling Theory with Experiment\relax
\mciteBstWouldAddEndPuncttrue
\mciteSetBstMidEndSepPunct{\mcitedefaultmidpunct}
{\mcitedefaultendpunct}{\mcitedefaultseppunct}\relax
\EndOfBibitem
\bibitem[Reuter(2016)]{reuter2016ab}
Reuter,~K. Ab initio thermodynamics and first-principles microkinetics for
  surface catalysis. \emph{Catalysis Letters} \textbf{2016}, \emph{146},
  541--563\relax
\mciteBstWouldAddEndPuncttrue
\mciteSetBstMidEndSepPunct{\mcitedefaultmidpunct}
{\mcitedefaultendpunct}{\mcitedefaultseppunct}\relax
\EndOfBibitem
\bibitem[Scheffler(1988)]{scheffler1988thermodynamic}
Scheffler,~M. In \emph{Physics of Solid Surfaces 1987}; Koukal,~J., Ed.;
  Studies in Surface Science and Catalysis; Elsevier, 1988; Vol.~40; pp
  115--122\relax
\mciteBstWouldAddEndPuncttrue
\mciteSetBstMidEndSepPunct{\mcitedefaultmidpunct}
{\mcitedefaultendpunct}{\mcitedefaultseppunct}\relax
\EndOfBibitem
\bibitem[Reuter and Scheffler(2001)Reuter, and
  Scheffler]{reuter2001composition}
Reuter,~K.; Scheffler,~M. Composition, structure, and stability of
  ${\mathrm{RuO}}_{2}(110)$ as a function of oxygen pressure. \emph{Physical
  Review B} \textbf{2001}, \emph{65}, 035406\relax
\mciteBstWouldAddEndPuncttrue
\mciteSetBstMidEndSepPunct{\mcitedefaultmidpunct}
{\mcitedefaultendpunct}{\mcitedefaultseppunct}\relax
\EndOfBibitem
\bibitem[Rogal and Reuter(2007)Rogal, and Reuter]{rogal2006ab}
Rogal,~J.; Reuter,~K. Ab initio atomistic thermodynamics for surfaces: A
  primer. \emph{Experiment, Modeling and Simulation of Gas-Surface Interactions
  for Reactive flows in Hypersonic Flights} \textbf{2007}, \emph{14},
  2--1\relax
\mciteBstWouldAddEndPuncttrue
\mciteSetBstMidEndSepPunct{\mcitedefaultmidpunct}
{\mcitedefaultendpunct}{\mcitedefaultseppunct}\relax
\EndOfBibitem
\bibitem[Tiwari \latin{et~al.}(2020)Tiwari, Heenen, Bjørnlund, Hochfilzer,
  Chan, and Horch]{tiwari2020electrochemical}
Tiwari,~A.; Heenen,~H.~H.; Bjørnlund,~A.~S.; Hochfilzer,~D.; Chan,~K.;
  Horch,~S. Electrochemical Oxidation of CO on Cu Single Crystals under
  Alkaline Conditions. \emph{ACS Energy Letters} \textbf{2020}, \emph{5},
  3437--3442\relax
\mciteBstWouldAddEndPuncttrue
\mciteSetBstMidEndSepPunct{\mcitedefaultmidpunct}
{\mcitedefaultendpunct}{\mcitedefaultseppunct}\relax
\EndOfBibitem
\bibitem[Tiwari \latin{et~al.}(2020)Tiwari, Heenen, Bjørnlund, Maagaard, Cho,
  Chorkendorff, Kristoffersen, Chan, and Horch]{tiwari2020fingerprint}
Tiwari,~A.; Heenen,~H.~H.; Bjørnlund,~A.~S.; Maagaard,~T.; Cho,~E.;
  Chorkendorff,~I.; Kristoffersen,~H.~H.; Chan,~K.; Horch,~S. Fingerprint
  Voltammograms of Copper Single Crystals under Alkaline Conditions: A
  Fundamental Mechanistic Analysis. \emph{The Journal of Physical Chemistry
  Letters} \textbf{2020}, \emph{11}, 1450--1455, PMID: 32022563\relax
\mciteBstWouldAddEndPuncttrue
\mciteSetBstMidEndSepPunct{\mcitedefaultmidpunct}
{\mcitedefaultendpunct}{\mcitedefaultseppunct}\relax
\EndOfBibitem
\bibitem[Gro{\ss} and Sakong(2022)Gro{\ss}, and Sakong]{gross2022ab}
Gro{\ss},~A.; Sakong,~S. Ab Initio Simulations of Water/Metal Interfaces.
  \emph{Chemical Reviews} \textbf{2022}, \emph{122}, 10746--10776, PMID:
  35100505\relax
\mciteBstWouldAddEndPuncttrue
\mciteSetBstMidEndSepPunct{\mcitedefaultmidpunct}
{\mcitedefaultendpunct}{\mcitedefaultseppunct}\relax
\EndOfBibitem
\bibitem[Ringe \latin{et~al.}(2022)Ringe, H\"{o}rmann, Oberhofer, and
  Reuter]{ringe2021implicit}
Ringe,~S.; H\"{o}rmann,~N.~G.; Oberhofer,~H.; Reuter,~K. Implicit Solvation
  Methods for Catalysis at Electrified Interfaces. \emph{Chemical Reviews}
  \textbf{2022}, \emph{122}, 10777--10820, PMID: 34928131\relax
\mciteBstWouldAddEndPuncttrue
\mciteSetBstMidEndSepPunct{\mcitedefaultmidpunct}
{\mcitedefaultendpunct}{\mcitedefaultseppunct}\relax
\EndOfBibitem
\bibitem[Dattila \latin{et~al.}(2022)Dattila, Seemakurthi, Zhou, and
  L\'{o}pez]{dattila2022modeling}
Dattila,~F.; Seemakurthi,~R.~R.; Zhou,~Y.; L\'{o}pez,~N. Modeling Operando
  Electrochemical CO2 Reduction. \emph{Chemical Reviews} \textbf{2022},
  \emph{122}, 11085--11130, PMID: 35476402\relax
\mciteBstWouldAddEndPuncttrue
\mciteSetBstMidEndSepPunct{\mcitedefaultmidpunct}
{\mcitedefaultendpunct}{\mcitedefaultseppunct}\relax
\EndOfBibitem
\bibitem[Nitopi \latin{et~al.}(2019)Nitopi, Bertheussen, Scott, Liu, Engstfeld,
  Horch, Seger, Stephens, Chan, Hahn, Nørskov, Jaramillo, and
  Chorkendorff]{nitopi2019progress}
Nitopi,~S.; Bertheussen,~E.; Scott,~S.~B.; Liu,~X.; Engstfeld,~A.~K.;
  Horch,~S.; Seger,~B.; Stephens,~I. E.~L.; Chan,~K.; Hahn,~C.;
  Nørskov,~J.~K.; Jaramillo,~T.~F.; Chorkendorff,~I. Progress and Perspectives
  of Electrochemical CO2 Reduction on Copper in Aqueous Electrolyte.
  \emph{Chemical Reviews} \textbf{2019}, \emph{119}, 7610--7672, PMID:
  31117420\relax
\mciteBstWouldAddEndPuncttrue
\mciteSetBstMidEndSepPunct{\mcitedefaultmidpunct}
{\mcitedefaultendpunct}{\mcitedefaultseppunct}\relax
\EndOfBibitem
\bibitem[Jordan \latin{et~al.}(1958)Jordan, Javick, and
  Ranz]{jordan1958hydrodynamic}
Jordan,~J.; Javick,~R.~A.; Ranz,~W.~E. Hydrodynamic Voltammetry at Solid
  Indicator Electrodes. \emph{Journal of the American Chemical Society}
  \textbf{1958}, \emph{80}, 3846--3852\relax
\mciteBstWouldAddEndPuncttrue
\mciteSetBstMidEndSepPunct{\mcitedefaultmidpunct}
{\mcitedefaultendpunct}{\mcitedefaultseppunct}\relax
\EndOfBibitem
\bibitem[Mitchell \latin{et~al.}(2001)Mitchell, Brown, and
  Rikvold]{Mitchell2001_Static_and_dynamic_MC}
Mitchell,~S.; Brown,~G.; Rikvold,~P. Static and dynamic Monte Carlo simulations
  of Br electrodeposition on Ag(100). \emph{Surface Science} \textbf{2001},
  \emph{471}, 125--142\relax
\mciteBstWouldAddEndPuncttrue
\mciteSetBstMidEndSepPunct{\mcitedefaultmidpunct}
{\mcitedefaultendpunct}{\mcitedefaultseppunct}\relax
\EndOfBibitem
\bibitem[H\"ormann and Gro\ss{}(2019)H\"ormann, and
  Gro\ss{}]{hoermann2019phase}
H\"ormann,~N.~G.; Gro\ss{},~A. Phase field parameters for battery compounds
  from first-principles calculations. \emph{Physical Review Materials}
  \textbf{2019}, \emph{3}, 055401\relax
\mciteBstWouldAddEndPuncttrue
\mciteSetBstMidEndSepPunct{\mcitedefaultmidpunct}
{\mcitedefaultendpunct}{\mcitedefaultseppunct}\relax
\EndOfBibitem
\bibitem[{\AA}ngqvist \latin{et~al.}(2019){\AA}ngqvist, Mu{\~n}oz, Rahm,
  Fransson, Durniak, Rozyczko, Rod, and Erhart]{aangqvist2019icet}
{\AA}ngqvist,~M.; Mu{\~n}oz,~W.~A.; Rahm,~J.~M.; Fransson,~E.; Durniak,~C.;
  Rozyczko,~P.; Rod,~T.~H.; Erhart,~P. ICET – A Python Library for
  Constructing and Sampling Alloy Cluster Expansions. \emph{Advanced Theory and
  Simulations} \textbf{2019}, \emph{2}, 1900015\relax
\mciteBstWouldAddEndPuncttrue
\mciteSetBstMidEndSepPunct{\mcitedefaultmidpunct}
{\mcitedefaultendpunct}{\mcitedefaultseppunct}\relax
\EndOfBibitem
\bibitem[Weitzner and Dabo(2017)Weitzner, and Dabo]{weitzner2017voltage}
Weitzner,~S.~E.; Dabo,~I. Voltage-dependent cluster expansion for electrified
  solid-liquid interfaces: Application to the electrochemical deposition of
  transition metals. \emph{Physical Review B} \textbf{2017}, \emph{96},
  205134\relax
\mciteBstWouldAddEndPuncttrue
\mciteSetBstMidEndSepPunct{\mcitedefaultmidpunct}
{\mcitedefaultendpunct}{\mcitedefaultseppunct}\relax
\EndOfBibitem
\bibitem[McCrum and Janik(2016)McCrum, and Janik]{mccrum2016ph}
McCrum,~I.~T.; Janik,~M.~J. pH and Alkali Cation Effects on the Pt Cyclic
  Voltammogram Explained Using Density Functional Theory. \emph{The Journal of
  Physical Chemistry C} \textbf{2016}, \emph{120}, 457--471\relax
\mciteBstWouldAddEndPuncttrue
\mciteSetBstMidEndSepPunct{\mcitedefaultmidpunct}
{\mcitedefaultendpunct}{\mcitedefaultseppunct}\relax
\EndOfBibitem
\bibitem[McCrum and Janik(2016)McCrum, and Janik]{mccrum2016first}
McCrum,~I.~T.; Janik,~M.~J. First Principles Simulations of Cyclic
  Voltammograms on Stepped Pt(553) and Pt(533) Electrode Surfaces.
  \emph{ChemElectroChem} \textbf{2016}, \emph{3}, 1609--1617\relax
\mciteBstWouldAddEndPuncttrue
\mciteSetBstMidEndSepPunct{\mcitedefaultmidpunct}
{\mcitedefaultendpunct}{\mcitedefaultseppunct}\relax
\EndOfBibitem
\bibitem[Heenen \latin{et~al.}(2020)Heenen, Gauthier, Kristoffersen, Ludwig,
  and Chan]{heenen2020solvation}
Heenen,~H.~H.; Gauthier,~J.~A.; Kristoffersen,~H.~H.; Ludwig,~T.; Chan,~K.
  {Solvation at metal/water interfaces: An ab initio molecular dynamics
  benchmark of common computational approaches}. \emph{The Journal of Chemical
  Physics} \textbf{2020}, \emph{152}, 144703\relax
\mciteBstWouldAddEndPuncttrue
\mciteSetBstMidEndSepPunct{\mcitedefaultmidpunct}
{\mcitedefaultendpunct}{\mcitedefaultseppunct}\relax
\EndOfBibitem
\bibitem[Yarnell \latin{et~al.}(1973)Yarnell, Katz, Wenzel, and
  Koenig]{yarnell1973structure}
Yarnell,~J.~L.; Katz,~M.~J.; Wenzel,~R.~G.; Koenig,~S.~H. Structure Factor and
  Radial Distribution Function for Liquid Argon at 85
  \ifmmode^\circ\else\textdegree\fi{}K. \emph{Physical Review A} \textbf{1973},
  \emph{7}, 2130--2144\relax
\mciteBstWouldAddEndPuncttrue
\mciteSetBstMidEndSepPunct{\mcitedefaultmidpunct}
{\mcitedefaultendpunct}{\mcitedefaultseppunct}\relax
\EndOfBibitem
\bibitem[Zhang \latin{et~al.}(2007)Zhang, Blum, and Reuter]{zhang2007accuracy}
Zhang,~Y.; Blum,~V.; Reuter,~K. Accuracy of first-principles lateral
  interactions: Oxygen at Pd(100). \emph{Physical Review B} \textbf{2007},
  \emph{75}, 235406\relax
\mciteBstWouldAddEndPuncttrue
\mciteSetBstMidEndSepPunct{\mcitedefaultmidpunct}
{\mcitedefaultendpunct}{\mcitedefaultseppunct}\relax
\EndOfBibitem
\bibitem[Taylor \latin{et~al.}(1985)Taylor, Williams, Park, Bartelt, and
  Einstein]{taylor1985two}
Taylor,~D.~E.; Williams,~E.~D.; Park,~R.~L.; Bartelt,~N.~C.; Einstein,~T.~L.
  Two-dimensional ordering of chlorine on Ag(100). \emph{Physical Review B}
  \textbf{1985}, \emph{32}, 4653--4659\relax
\mciteBstWouldAddEndPuncttrue
\mciteSetBstMidEndSepPunct{\mcitedefaultmidpunct}
{\mcitedefaultendpunct}{\mcitedefaultseppunct}\relax
\EndOfBibitem
\end{mcitethebibliography}

\end{document}